\documentclass[11pt,preprint,aps,amsmath,amssymb,prd]{revtex4}

\usepackage{graphicx}
\usepackage{bm}
\usepackage{color}

\newcommand{\modif}{}

\begin{document}

\def\be{\begin{equation}}
\def\ee{\end{equation}}
\def\d{\mbox{\rm d}}

\title{Cosmological structure formation with negative mass}

\author{Giovanni Manfredi}
\email{giovanni.manfredi@ipcms.unistra.fr}
\affiliation{Universit\'e de Strasbourg, CNRS, Institut de Physique et Chimie des Mat\'eriaux de Strasbourg, UMR 7504, F-67000 Strasbourg, France}
\author{Jean-Louis Rouet}
\affiliation{Universit\'e d'Orl\'eans, CNRS/INSU, BRGM, ISTO, UMR7327, F-45071 Orl\'eans, France}
\author{Bruce Miller}
\affiliation{Department of Physics and Astronomy, Texas Christian University, Fort Worth, TX 76129}
\author{Gabriel Chardin}
\affiliation{Centre National de la Recherche Scientifique, Rue Michel-Ange, F-75016 Paris, France}

\date{\today}

\begin{abstract}
We construct a family of models with negative gravitational mass in the context of Newtonian  gravity. We focus in particular on a model that reproduces the features of the so-called Dirac-Milne universe, a matter-antimatter symmetric universe that was recently proposed as an alternative cosmological scenario [A. Benoit-L{\'e}vy and G. Chardin, A\&A {\bf 537}, A78 (2012)].
We perform one-dimensional N-body simulations of these negative-mass models for an expanding universe and study the associated formation of gravitational structures. The similarities and differences with the standard cosmological model are highlighted and discussed.
\end{abstract}

\maketitle

\section{Introduction}\label{sec:intro}

It may seem unlikely to study cosmological structure formation using hypothetical negative mass objects. On the other hand, we cannot but note the strangeness of the Standard Cosmological Model which, although impressively concordant on primordial nucleosynthesis, cosmic microwave background (CMB), baryon acoustic oscillations (BAO) and type-1a supernovae (SN1a) luminosity distance, features a very strange composition, with dark matter and dark energy, two unidentified components, supposedly representing approximately $96\%$ of the universe content.

In the standard ($\Lambda \rm CDM$) model, it is also difficult to understand the different periods of dominance and insignificance of the various components of the Universe: on one hand, dark energy is remarkably insignificant in the very early stages of the universe, representing less than $10^{-120}$ of the energy density at that epoch, and remaining  almost completely negligible until recently. On the other, dark energy, if it exists, is supposed to represent at present about $70\%$ of the universe energy density, and will become strongly predominant in the future. Similarly, matter appears as an insignificant component in the radiation dominated universe from the quark-gluon plasma transition down to temperatures of a few electron-Volts and, after a period of matter domination, will soon become again insignificant.
We should therefore be ready to examine the proposition that our description of the universe following the Standard Cosmological Model, although it may represent a fair adjustment to the experimental data, has no connection with the actual composition of the universe.

Looking for descriptions alternative to the Standard Cosmological Model, several authors have noted that our universe is very close to a ``coasting" universe, i.e., a universe that neither decelerates nor accelerates, and therefore behaves, at large scales ($\gtrsim 100 \rm Mpc$), as a gravitationally empty, or Milne \cite{Milne}, universe. For example, Nielsen, Guffanti and Sarkar (NGS) \cite{NGS} argue that the present SN1a data, twenty years after the discovery of repulsive gravity, is still unable to demonstrate convincingly the acceleration of the universe expansion rate. Similarly, by expanding on the NGS analysis, Blanchard and collaborators \cite{Tutu2017} have shown that cosmic acceleration by all ``local" cosmological probes (at $z < 3$, where $z$ is the cosmological redshift) is not statistically compelling, while recently the SN1a data of the Hubble Space Telescope \cite{Riess2017} show that, at the very least, the Milne universe appears as an excellent fit to these SN1a data. Also, it is well known that the age of our $\Lambda \rm CDM$ universe is nearly equal to the age of a Milne universe (see \cite{Benoitlevy} and references therein).

These are motivations to look for the possibility that negative mass particles exist in the universe, which would incorporate equal amounts of positive and negative mass particles, thus appearing gravitationally empty. While negative mass solutions seem to violate blatantly the sacrosanct energy conditions, it should be noted that counterexamples have been found to basically all the expressions of energy conditions (for a review, see \cite{Barcelo2002}). Indeed, dark energy itself, with a pressure equal to the negative of its energy density, violates the strong energy condition. On the other hand, recently, Paranjape and collaborators \cite{Mbarek2014} have explicitly constructed perfectly physically acceptable negative-mass ``bubbles" sitting in an expanding Einstein-de Sitter universe.

A further motivation to study a universe featuring equal amounts of positive and negative mass particles is provided by the so-called ``Dirac-Milne" universe \cite{Benoitlevy}, i.e., a symmetric matter-antimatter universe where antimatter particles have a negative gravitational mass, which is remarkably concordant without any adjustable parameter. Matter-antimatter universes have been studied by Omn\`es and his group in the late sixties \cite{Omnes1972}, and later by Cohen, de Rujula and Glashow \cite{Cohen1998}, and it would seem that they are excluded by observation: annihilation at the matter-antimatter domain interfaces produces diffuse gamma-ray flux in the tens of MeV range that contradict existing observational constraints, unless the size of matter and antimatter domains exceeds significantly the gigaparsec scale. However, these two studies suppose, quite understandably, that antimatter respects the Equivalence Principle (EP), which lies at the heart of general relativity.

Is there a simple and natural modification of the EP for antimatter that avoids the annihilation constraint? To answer this question, we note that there exists a physical system, the electron-hole gas in a semiconductor, where pseudo-particles and their antiparticles avoid annihilation by creating a charge-free region -- i.e., a depletion zone -- under either an electric or gravitational field. In this system, electrons gravitate and holes antigravitate \cite{Tsidi1975}. The electron-hole system in a semiconductor provides a physical implementation of Dirac's original idea of the positron as an electron hole in a Dirac sea. A natural redefinition of the EP using two types of negative and positive mass carriers (by analogy with negative and positive charge carriers) was first proposed by Piran \cite{Piran1997} following the simulations by Dubinski et al. \cite{Dubinski1993}, where they considered underdense regions as negative mass regions, with counterintuitive repulsive behavior.
The fact that underdense regions lead to nonlinear structures of progressively larger size, providing discriminating cosmological tests, has been studied in particular by Sheth and collaborators \cite{sheth2004,colberg2005voids}.

To date, there is no clear-cut direct experimental evidence on the gravitational behavior of antimatter. But this situation may soon change, with several precision experiments currently undertaken at CERN to test the gravitational response of neutral antihydrogen atoms. In particular, the first results of the Gbar \cite{Indelicato2014}, ALPHA-g \cite{alpha-g}, and AEgIS \cite{AEGIS} collaborations should be available before the end of this decade, and any deviation from perfect matter-antimatter symmetry will have profound cosmological implications.

In the present work, we first study the various sign combinations of inertial, active and passive gravitational mass that could be used to define negative mass particles in the context of Newtonian gravity. Remarkably, we find that the system corresponding to the analog of an electron-hole system, motivating the Dirac-Milne universe, cannot be represented by any combination of such Newtonian masses, whatever their signs. Instead, the Dirac-Milne scenario must be represented by a matrix Poisson equation that couples unconventionally the two particle species.

Subsequently, we investigate gravitational structure formation in such negative-mass universes by means of numerical simulations of an expanding one-dimensional universe, focusing in particular on the Dirac-Milne scenario.
Significantly, in this last case, simulations show that structures are produced at relatively early epochs and remain ``frozen" in the subsequent history of the universe, without requiring either inflation or dark matter ingredients.

\section{Newtonian cosmology with negative mass}\label{sec:negmass}
The possibility that particles with negative mass exist has long been considered (and sometimes dismissed) in the past \cite{Price_ajp1993, Mannheim_foundphys2000,Hammond_ejp2015}, starting from the seminal work of the late Hermann Bondi \cite{Bondi}. Negative mass is not necessarily incompatible with fundamental physical laws, such as the equivalence principle, the conservation of energy and momentum, and Newton's third (action/reaction) law. Nevertheless, it does give rise to downright unusual features \cite{Hammond_ejp2015}, such as the runaway acceleration of two opposite masses placed near each other, or the fact that, although a positive and a negative mass will individually fall in a gravitational field, the bound system of these two objects will levitate and polarize, with the negative mass levitating above the positive one \cite{Price_ajp1993}.
In this work, our attention will be focussed on the cosmological consequences of matter that displays a negative mass. This question was raised in the context of Milgrom's Modified Newtonian Dynamics (MOND) \cite{Milgrom_1986}. Cosmological simulations using negative Bondi masses were also reported in a recent study \cite{Farnes2017}.

It should be added that negative {\em effective} mass is a rather old and uncontroversial topic in such disparate fields as materials science \cite{negmass_acousticmetam}, semiconductor electronics \cite{negmass_thz}, and cold atomic gases \cite{negmass_bec}. Usually, it means that a certain wave dispersion relation contains a negative term that can be interpreted, by analogy, as a ``negative mass".

In the remainder of this section, we review different models of negative mass in the context of Newtonian gravity. Some of these models have been considered in the past, such as the Bondi masses mentioned above or the standard antigravity scenario whereby unlike masses repel and like masses attract each other. But other cases are possible if one relaxes the hypothesis that Newtonian gravity is fully described by a single gravitational potential, and considers instead {\em two} potentials. This approach, described in more detail in Sec. \ref{sec:diracmilne}, can be viewed as the nonrelativistic limit of a bimetric theory of gravity \cite{Hohmann_PRD2009,Hohmann_PRD2010}. In this perspective, it is possible to derive a theory that possesses the features required by the Dirac-Milne universe proposed recently by Benoit-L\'evy and Chardin \cite{Benoitlevy}.
Within the Dirac-Milne cosmology, it is ordinary antimatter that carries a negative mass. In this spirit, we will often refer to particles with positive and negative mass respectively as ``matter" and ``antimatter" although, generally speaking, negative mass particles need not be identified with antimatter: they are just another kind of matter that responds differently to gravitational fields.

\subsection{Negative active or passive gravitational masses}\label{sec:activepassive}
Let us first consider the cases that can be obtained easily by posing different signs for the various types of masses. As is customary, we distinguish between the active $m_a$ and passive $m_p$ gravitational masses, in addition to the inertial mass $m_i$.
The active gravitational mass is the one that appears on the right-hand side of Poisson's equation:
\be
\Delta \phi = 4\pi G \rho =  4\pi G m_a n,
\ee
where $n$ is the number density, $\rho$ the mass density, and $\phi$ the gravitational potential.
The passive gravitational mass relates the force to the gravitational field through: ${\bm F} = -m_p \nabla\phi$.
Finally, the inertial mass gives the relation between velocity and momentum: ${\bm p} = m_i \dot {\bm r}$ (the dot denotes differentiation with respect to time). Therefore Newton's second law of motion $\dot {\bm p}={\bm F}$ becomes
\be
\ddot {\bm r} = -(m_p/m_i) \nabla\phi  \label{eq:newton}.
\ee
The equivalence principle (EP) demands that the inertial mass be identical to the  passive gravitational mass, so that a gravitational field becomes equivalent to an acceleration. But the EP says nothing about the active gravitational mass, which may in principle be different.
On the other hand, Newton's third law (the action/reaction principle) requires that the passive and active gravitational masses be the same; otherwise the total force acting on an isolated system of interacting particles would be nonzero, and momentum conservation would be violated.

Since only the ratio $m_p/m_i$ appears in this Newtonian context, we will assume that the inertial mass is positive and consider the different signs for the active and passive gravitational masses.
All possible cases are summarized in Table I.

\begin{table*}[ht!]
\begin{tabular}{|c|c|c|c|c|c|}
\hline
  & Case & Name& Active grav. mass & Passive grav. mass  & Inertial mass \\
\hline
\hline
Matter &A & Standard &+ &  +  & +  \\
\hline
 & B & Antiplasma & $-$ & $-$ & + \\
Antimatter & C & Bondi & $-$ & + & + \\
 & D & Antiinertia & $+$ & $-$ & $+$ \\
\hline
\end{tabular}
\caption{Signs of the three types of mass for the four cases considered in the text. The absolute values of the masses are supposed to be the same. As only the ratio $m_p/m_i$ is relevant, we assume for all cases $m_i>0$, with no loss of generality.}
\end{table*}

Case A is the standard one, with all three masses being positive.
For the other cases, we consider a two-component system composed of both matter (type-A particles in Table I) and antimatter (types B, C, or D).
Case B is the ``antiplasma" scenario, so named because it is the analog of a two-component plasma (made of electrons and ions interacting via the Coulomb force) but with opposite sign for the interaction: like masses attract, whereas unlike masses repel each other. Note that case B does not respect the EP ($m_p \neq m_i$).
Case C is that of the so-called Bondi masses \cite{Bondi}: positive masses attract everything; negative masses repel everything. This case respects the EP, because $m_p = m_i$, but gives rise to some surprising features, like the runaway acceleration of a pair of positive and negative masses. Indeed, if a positive and a negative mass are placed near each other, the positive mass will attract the negative mass and the negative mass will repel the positive one: the net result is that both masses accelerate without bound.
As only the ratio $m_p/m_i$ is relevant, the case of Bondi masses may also be viewed as a case where {\em all} three masses are negative. This is actually necessary to ensure kinetic energy and momentum conservation during the above-mentioned runaway acceleration.

The case D was never considered before, at least to our knowledge.
In this case, both matter and antimatter contribute in the same way to the gravitational field (because their active gravitational masses are both positive). However, while matter is attracted by the field, antimatter is repelled, because it has negative passive gravitational mass. We termed this case ``antiinertia" because it can also be viewed (by changing the signs of both $m_p$ and $m_i$) as a case where both gravitational masses are positive while the inertial mass is negative.

Finally, let us note that of the four possible cases of Table I, only A (all masses positive) and C (all masses negative) satisfy at the same time the action/reaction principle and the EP.

In the rest of this work, we will deal with two-component systems made of one positive-mass species (case A, ``matter") and a negative-mass species (``antimatter"). For the latter, we will only consider cases B (antiplasma) and C (Bondi). In the Newtonian limit and in the absence of two-body correlations (collisionless approximation), such a system is governed by the following Vlasov-Poisson equations:
\begin{eqnarray}
\frac{\partial f_{\pm}}{\partial t} &+&
{\bm v}\cdot\nabla f_{\pm} - \,\frac{m_p^{\pm}}{m_i^{\pm}} {\nabla \phi} \cdot\nabla_{\bm v} f_{\pm} = 0, \label{vlasov-poisson} \\
\Delta\phi &=& 4\pi G (m_a^{+} n_{+} + m_a^{-}n_{-}),
\end{eqnarray}
where $f_{\pm}({\bm r},{\bm v},t)$ are the probability densities (distribution functions) for matter ($+$) and antimatter ($-$) in the phase space, and $n_{\pm}=\int f_{\pm} d{\bm v}$ are the number densities. All masses have the same absolute value ($|m_a|=|m_p|=|m_i|=m$) and their signs are given in Table I.

\subsection{Dirac-Milne scenario}\label{sec:diracmilne}

The idea of a matter-antimatter symmetric universe has already been considered in the past \cite{Cohen1998,Ramani76}, but was excluded because annihilations at the frontiers of matter and antimatter domains would generate a diffuse gamma ray emission that would be in contradiction with observational data (unless the matter domain we inhabit is virtually the entire visible universe).

Recently, Benoit-Levy and Chardin \cite{Benoitlevy} considered an alternative matter-antimatter symmetric cosmological scenario, where the gravitational interaction between matter and antimatter is repulsive. The latter property is a necessary condition to avoid contact and annihilation between matter and antimatter domains after cosmological recombination.
Such matter-antimatter repulsion could in principle be achieved by cases B (antiplasma) and C (Bondi) in Table I. However, the antiplasma scenario does not respect the EP and implies the existence of antimatter gravitational structures resulting from antimatter-antimatter attraction, again leading to unwanted annihilation at the boundaries between matter and antimatter regions.
The Bondi case would require a negative inertial mass to ensure energy and momentum conservation, and displays peculiar and counterintuitive features such as the runaway acceleration.

In order to reproduce the Dirac-Milne universe postulated by Benoit-Levy and Chardin \cite{Benoitlevy}, we would need a scenario in which matter forms self-gravitating structures such as galaxies and clusters of galaxies, whereas antimatter constitutes a diffuse low-density background. This can be achieved if the following conditions are satisfied: (i) matter attracts matter, (ii) antimatter repels antimatter, and (iii) antimatter and matter repel each other -- these conditions are summarized in Table II.
The difference with respect to the antiplasma case is that for the Dirac-Milne scenario negative masses repel each other, whereas in the antiplasma case they attract.
Also, contrarily to the Bondi case, matter repels antimatter in the Dirac-Milne scenario.

In the Dirac-Milne scenario, matter forms gravitational structures, whereas antimatter, being repelled by everything (including itself), tends to spread across all the available space. Such spread is almost uniform, but not quite: since matter repels antimatter, the latter is expelled from matter-dominated overdense regions (galaxies) and forms a low-density almost homogeneous background distributed over the underdense regions in between matter's gravitational structures. This behavior precludes the occurrence of any significant annihilation, in accordance with observation.
In the following analyses and computer simulations, we will often make the simplifying hypothesis that antimatter constitutes a low-density homogeneous background uniformly distributed everywhere in space. This approximation is justified for the study of gravitational structure formation, as overdense regions are very much dominated by matter anyway.

\begin{table*}[ht!]
\begin{tabular}{|c|c|c|c|}
\hline \hline
Type of matter & Type of matter & Interaction\\
\hline
\hline
$+$ & $+$ & Attraction\\
\hline
$-$ & $-$ & Repulsion \\
\hline
$-$ & $+$ & Repulsion \\
\hline
$+$ & $-$ & Repulsion \\
\hline
\end{tabular}
\caption{Interactions between matter ($+$) and antimatter ($-$) particles in the Dirac-Milne universe.}
\end{table*}

However, the Dirac-Milne scenario described in the above paragraph does not have a proper Newtonian limit, in the sense that no combination of the signs of the three masses (as in Table I) can reproduce it. The Dirac-Milne scenario can only be accounted for by {\em two} gravitational potentials that obey two distinct Poisson equations. In the collisionless approximation, this leads to the following Vlasov-Poisson system:
\begin{eqnarray}
\frac{\partial f_{\pm}}{\partial t} &+&
{\bm v}\cdot\nabla f_{\pm} - \nabla \phi_{\pm} \cdot\nabla_{\bm v} f_{\pm} = 0, \label{vlasov_diracmilne}  \\
\Delta\phi_{+} &=& 4\pi G m(+n_{+} - n_{-}), \label{poiss_diracmilne1}\\
\Delta\phi_{-} &=& 4\pi G m(-n_{+} - n_{-}) \label{poiss_diracmilne2}
\end{eqnarray}
The above equations cannot be reduced to one of the cases summarized in Table I. This is because there are actually {\em two} active gravitational masses for matter, positive in Eq. \eqref{poiss_diracmilne1} and negative in Eq. \eqref{poiss_diracmilne2} (for antimatter, the active gravitational mass is always negative).
It can be proven that the above equations conserve the total momentum
\[
{\bm P}_{\rm tot}=\int\int m{\bm v}\,\left(f_{+}+f_{-}\right)d{\bm r}d{\bm v}
\]
and the total energy
\[
E_{\rm tot}={m \over 2}\int\int (f_+ + f_-)|{\bm v}|^2 d{\bm r}d{\bm v} +
 \frac{1}{8\pi G}\int \left(\frac{|\nabla\phi_{-}|^2 -  |\nabla\phi_{+}|^2}{2} +
 \nabla\phi_{-}\cdot\nabla\phi_{+}\right) d{\bm r}.
 \]

In the remainder of this paper, we will elaborate on some of the consequences of three of the models described so far, namely the antiplasma, Bondi, and Dirac-Milne cases. In particular, we will be concerned with the problem of cosmological structure formation in an expanding universe.
But before doing that, we show in the next subsection that all the aforementioned scenarios can be described in the framework of a single general formalism.

\subsection{General formalism} \label{sec:genform}
The representation of Newtonian gravity with two Poisson equations, as in Eqs. \eqref{poiss_diracmilne1}-\eqref{poiss_diracmilne2}, is not entirely new. For instance, Hohmann and Wohlfart \cite{Hohmann_PRD2009} considered a general-relativistic theory in which two classes of particles, respectively with positive and negative mass ratios $m_p/m_i$, are described by different metric tensors.
They further proved a no-go theorem that rules out all bimetric theories where like particle attract and unlike particles repel each other, i.e., the analog of the antiplasma case that we described in Sec. \ref{sec:negmass}.
However, the Dirac-Milne scenario considered in the present work should not be affected by this theorem.
Hossenfelder \cite{Hossen_PRD2008} also developed a similar bimetric theory which putatively does not suffer from the restriction of the above-mentioned no-go theorem \cite{Hossen_arxiv2009}.

All the scenarios considered in the preceding sections can be represented in matrix form with the following Vlasov-Poisson equations
\begin{eqnarray}
\frac{\partial f_\pm}{\partial t} &+&
{\bm v}\cdot\nabla f_{\pm} - \nabla{\phi_\pm}\cdot\nabla_{\bm v} f_{\pm}= 0, \label{vlasov_matrix}\\
\Delta \Phi &=& 4\pi G m ~ \widehat{\textsf{M}} \, {\textsf n} , \label{poisson_matrix}
\end{eqnarray}
where
\be
\Phi = \begin{pmatrix}\phi_{+} \\ \phi_{-}  \end{pmatrix},
\quad
{\textsf n} = \begin{pmatrix}n_{+} \\ n_{-}  \end{pmatrix},
\quad
\widehat{\textsf{M}} = \begin{pmatrix}M_{++} & M_{+-}\\ M_{-+} & M_{--} \end{pmatrix}\, ,
\ee
and $M_{ij} = \pm 1$. Thus, $M_{++}$ indicates the effect of matter (+) on matter (+), $M_{+-}$ the effect of antimatter ($-$) on matter (+), $M_{-+}$ the effect of matter (+) on antimatter ($-$), and $M_{--}$ the effect of antimatter ($-$) on antimatter ($-$). For instance, $M_{++} = +1$ means that matter attracts matter, $M_{+-} = -1$ means that antimatter repels matter, $M_{--}=-1$ means that antimatter repels antimatter, etc.

The three cases described in Sec. \ref{sec:negmass} (antiplasma, Bondi, and antiinertia) are represented by the following matrices:
\be
\widehat{\textsf{M}}_{\rm ap} = \begin{pmatrix}1 & -1\\ -1 & 1 \end{pmatrix},
\quad
\widehat{\textsf{M}}_{\rm Bondi} = \begin{pmatrix}1 & -1\\ 1 & -1 \end{pmatrix},
\quad
\widehat{\textsf{M}}_{\rm ai} = \begin{pmatrix}1 & 1\\ -1 & -1 \end{pmatrix} .
\label{eq:matrix1}
\ee
Note that in all the above cases the matrix $\widehat{\textsf{M}}$ is non-invertible, i.e. $\det\widehat{\textsf{M}} =0$. This is expected, as such cases can be described by a single Poisson equation.

Let us now consider the most general case for the matrix $\widehat{\textsf{M}}$. One must always have $M_{++}=1$, because this is the standard behavior of ordinary matter (attraction). The other three elements of the matrix can be chosen freely, which leaves us with a total of $2^3=8$ possibilities. The special case where all elements are positive is trivial, as it corresponds to a situation where matter and antimatter are gravitationally indistinguishable. Then we have the three cases described by Eq. (\ref{eq:matrix1}).
This leaves us with another four cases that cannot simply be represented by a single Poisson equation. One of them is the Dirac-Milne case described in Sec. \ref{sec:diracmilne}, which yields the matrix:
\be
\widehat{\textsf{M}}_{\rm DM} = \begin{pmatrix}1 & -1\\ -1 & -1 \end{pmatrix}.
\label{eq:matrixdm}
\ee
The remaining three cases can be written as
\be
\widehat{\textsf{M}}_{1} = \begin{pmatrix}1 & 1\\ 1 & -1 \end{pmatrix},
\quad
\widehat{\textsf{M}}_{2} = \begin{pmatrix}1 & 1\\ -1 & 1 \end{pmatrix},
\quad
\widehat{\textsf{M}}_{3} = \begin{pmatrix}1 & -1\\ 1 & 1 \end{pmatrix}.
\label{eq:matrix2}
\ee
Note that in the cases of Eqs. (\ref{eq:matrixdm})-(\ref{eq:matrix2}) the matrix $\widehat{\textsf{M}}$ is invertible, i.e. $\det\widehat{\textsf{M}} \neq 0$, so that they cannot be reduced to one single Poisson equation.
All cases where the matrix is invertible should not be affected by the no-go theorem of Hohmann et al. \cite{Hohmann_PRD2009}.

It is easily shown that the matrix Poisson equation \eqref{poisson_matrix} is equivalent to the following Lagrangian:
\be
\mathcal{L(\phi_{+},\phi_{-})} = \frac{\nabla\Phi^T \cdot \nabla\Phi}{8\pi G} + m\,\Phi^T \widehat{\textsf{M}}\,\textsf{n},
\label{lagrangian}
\ee
where the superscript $T$ denotes the transpose matrix.

Some final considerations on the above models are in order here. Four of them have the unpleasant property that the interaction between unlike particles is not symmetric -- matter attracts antimatter but antimatter repels matter, or viceversa. This group of models include the Bondi, antiinertia, $\widehat{\textsf{M}}_{2}$, and $\widehat{\textsf{M}}_{3}$ cases. As a consequence, they all display the peculiar runaway acceleration effect that we described in Sec. \ref{sec:negmass} for the Bondi masses. This seems to be a good argument to rule out such models.

Of the remaining models, the antiplasma case leads to the formation of overdense gravitational structures for both matter and antimatter (``antigalaxies"), for which there is no observational evidence. The numerical simulation of this model (see Sec. \ref{sec:numres-antibondi}) also reveals some unobserved features, such as the creation of matter and antimatter streams with opposite velocities.

Finally, the case described by the matrix $\widehat{\textsf{M}}_{1}$ is similar to the Dirac-Milne scenario, with the significant difference that unlike masses attract each other. Again, this is bound to lead to many annihilation events because matter and antimatter would tend to cluster together.

All in all, the Dirac-Milne case is singled out as the most likely candidate to reproduce the features of the alternative cosmology proposed by Benoit-L\'evy and Chardin \cite{Benoitlevy}.

\section{Steady states and linear response analysis}\label{sec:linear}
It is interesting to note that, out of the eight possible cases mentioned in Sec. \ref{sec:genform}, spatially homogeneous steady states (i.e., with $n_{+} = n_{-} = \rm const.$ and $\phi_\pm =0$) can only exist for the antiplasma and Bondi cases. This is because they are the only cases for which positive and negative mass densities can compensate each other. This is not a problem {\em per se} as ordinary gravity itself does not admit steady state solutions.

{ For a Dirac-Milne universe, one can assume, to an excellent approximation, that antimatter constitutes a homogeneous dilute background with density $n_0$, as was discussed in Sec. \ref{sec:diracmilne}. In that case, one can drop the second Poisson equation \eqref{poiss_diracmilne2} and a steady state becomes indeed possible for $n_+=n_0$. This scenario will be justified on more physical grounds when we introduce comoving coordinates in an expanding frame of reference for cosmological applications, see Sec. \ref{sec:scaling}.

Hereafter, we present a linear response analysis of the Vlasov-Poisson equations for the three cases of interest: antiplasma, Bondi, and Dirac-Milne.
We expand the distribution functions $f_\pm({\bm r},{\bm v},t)=f_0({\bm v}) + \tilde f_{\pm}({\bm r},{\bm v},t)$, the densities $n_\pm=n_0 + \tilde n_{\pm}({\bm r},t)$, and the potentials $\phi_\pm({\bm r},t) = \phi_0+\tilde\phi_\pm({\bm r},t)$ (where the tilde denotes a small perturbation) and only retain first-order terms.
Then we Fourier analyze the perturbations in space and time, by writing for the density fluctuations:
$\tilde n_{\pm}({\bm r},t) = \tilde n_{\pm}\,\exp[i({\bm k}\cdot{\bm r}-\omega t)]$,
and analogous expressions for the other first-order quantities.

\subsection{Antiplasma}\label{sec:linear-anti}
From Eq. \eqref{vlasov-poisson} we get the relationship between the density perturbation and the potential perturbation:
\be
\tilde n_\pm =  \tilde \phi \int \frac{{\bm k}\cdot\nabla_{\bm v}f_0}{\omega-{\bm k}\cdot{\bm v}}\, d{\bm v}. \label{fluctuat}
\ee
Using Eq. \eqref{fluctuat} together with Poisson's equation in the antiplasma case, we arrive at the dispersion relation
\be
D(\omega,{\bm k}) \equiv 1-\frac{2\omega_{J0}^2}{k^2} \int\frac{{\bm k}\cdot\nabla_{\bm v}f_0}{\omega-{\bm k}\cdot{\bm v}}\, d{\bm v}= 0,
\ee
where $\omega_{J0} = (4\pi G m n_0)^{1/2}$ is the Jeans frequency. If ${\bm k}\cdot{\bm v} \ll \omega$, we can expand the denominator in the above integral
\[
\frac{1}{\omega -{\bm k}\cdot{\bm v}} \approx
\frac{1}{\omega}+\frac{{\bm k}\cdot{\bm v}}{\omega^2}+\frac{({\bm k}\cdot{\bm v})^2}{\omega^3}+\frac{({\bm k}\cdot{\bm v})^3}{\omega^4}
+ \dots \,.
\]
When the equilibrium distribution $f_0$ is a function of $v\equiv |{\bm v}|$ only, then the { dispersion relation} can be expressed in the following way:
\be
D(\omega, k) \approx 1 +\frac{2\omega_{J0}^2}{\omega^2}+ \frac{3
k^2 \langle v^2\rangle \omega_{J0}^2}{\omega^4},
\label{DL2}
\ee
where $ \langle v^2\rangle=\int v^2 f_0(v) dv$. Setting $D=0$ then yields:
\be
\omega^2 = -2\omega_{J0}^2 + 3k^2 \langle v^2\rangle,
\label{omega_anti}
\ee
and $\langle v^2\rangle = v_{th}^2 = k_BT/m$ for a Maxwell-Boltzmann distribution with temperature $T$. This expression was derived earlier \cite{Trigger_2015} using a fluid model.
Equation \eqref{omega_anti} is almost identical to the dispersion relation for self-gravitating systems in the presence of a repulsive background, displaying the usual Jeans instability for large wavelengths (small $k$). The extra factor 2 comes from having two mobile species, so that one should use the reduced mass $m_r = m_+^i m_{-}^i/(m_{+}^i + m_{-}^i) = m/2$.

\subsection{Bondi masses}\label{sec:linear-bondi}
In the Bondi case, the Vlasov-Poisson equations read as follows:
\begin{eqnarray}
\frac{\partial f_\pm}{\partial t} &+&
{\bm v}\cdot \nabla f_\pm -\nabla \phi \cdot \nabla_{\bm v} f_\pm = 0, \label{vlasov_bondi} \\
\Delta \phi &=& 4\pi G m  \left(n_{+} -n_{-}\right).
\end{eqnarray}
We note immediately that the Vlasov equations for $f_\pm$ are completely identical. It is therefore useful to define $g \equiv f_{+} + f_{-}$ and $h \equiv f_{+} - f_{-}$,
which obey the equations
\begin{eqnarray}
\frac{\partial g}{\partial t} &+&
{\bm v}\cdot \nabla g -\nabla \phi \cdot \nabla_{\bm v} g = 0, \label{eq:g}\\
\frac{\partial h}{\partial t} &+&
{\bm v}\cdot \nabla h -\nabla \phi \cdot \nabla_{\bm v} h = 0, \label{eq:h}\\
\Delta \phi &=&  4\pi Gm\int_{-\infty}^{\infty} h({\bm r},{\bm v},t)~ d {\bm v}. \label{eq:poisson_h}
\end{eqnarray}
We note that the equation for $g$ is decoupled from the equation for $h$. The system \eqref{eq:h}- \eqref{eq:poisson_h} evolves self-consistently independently of $g$, which is simply transported in the phase space like a passive scalar.
The system \eqref{eq:h}-\eqref{eq:poisson_h} looks identical to the equations of a single gravitational species evolving in standard Newtonian gravity, but there is an important difference: the distribution function $h$ can take both positive and negative values and in general will have zero mean.

To perform the linear analysis, we consider Eqs. \eqref{eq:h}-\eqref{eq:poisson_h} and expand all quantities to first order, remembering that at zeroth order $h_0=0$ and $\phi_0=0$. It follows that the first-order equations reduce to the trivial free transport equation:
\be
\frac{\partial \tilde h}{\partial t} + {\bm v}\cdot \nabla\tilde h =0.
\ee
There is no field-dependent term at first order and therefore no Jeans instability. Thus, nontrivial effects will only arise in the nonlinear regime.

\subsection{Dirac-Milne}\label{sec:linear-dm}
As mentioned earlier on, there is no homogeneous steady-state solution of Eqs. \eqref{vlasov_diracmilne}-\eqref{poiss_diracmilne2}, since antimatter acts repulsively with respect to both matter and antimatter. However, for the very same reason, antimatter tends to spread throughout space thus yielding an approximately uniform background with density $n_0$, as was discussed in Sec. \ref{sec:diracmilne}. In that case,
one can drop the second Poisson equation \eqref{poiss_diracmilne2} to obtain the following Vlasov-Poisson system for positive-mass matter
\begin{eqnarray}
\frac{\partial f_{+}}{\partial t} &+&
{\bm v}\cdot\nabla f_{+} - \nabla \phi_{+} \cdot\nabla_{\bm v} f_{+} = 0, \label{vlasov_diracmilne+}  \\
\Delta\phi_{+} &=& 4\pi G m(n_{+} - n_{0}). \label{poiss_diracmilne}
\end{eqnarray}
These are identical to the equations of ordinary gravity in the presence of a neutralizing background. Following the same procedure as done above for the antiplasma case, we obtain the usual Jeans dispersion relation
\be
\omega^2 = -\omega_{J0}^2 + 3k^2 \langle v^2\rangle,
\label{omega_dm}
\ee
which is identical to Eq. \eqref{omega_anti} except for the missing factor 2 in front of $\omega_{J0}$.

\section{Comoving coordinates and scaling}\label{sec:scaling}

In this section, we introduce the comoving coordinates used to study an expanding universe in Newtonian gravity and adapt them to the case of a matter-antimatter universe with both attractive and repulsive gravity.
Let us consider an expanding distribution of matter with spherical symmetry.
Its gravitational field has only one component $E_r(r,t)$ that depends on time and on a single spatial variable $r$. This type of system was studied extensively in the past \cite{Rouet1,Rouet2,MRexp,MR2010,miller2010ewald,Manfredi_PRE2016,quintic,Joyce2011,Benhaiem11032013}.

The equation of motion reads as (we ignore, for simplicity of notation, the signs of the masses; these will be reinstated later):
\be
\frac{d^2 r}{dt^2} = E_r(r,t),
\label{motion}
\ee
where $E_r=-\partial_r \phi$ is the gravitational field.
We consider an expanding universe with scaling factor $a(t)$ and transform space and time as follows:
\begin{eqnarray}
r &=& a(t) \hat r,  \label{scaling_r} \\
dt &=& b^2(t) d\hat t,  \label{scaling_t}
\end{eqnarray}
where comoving coordinates are denoted by an overcaret. Note that we also introduced a scaled time $\hat t$, which defines a new time-dependent ``clock".
The velocity transforms as
\be
\frac{\d r}{\d t} = \frac{a}{b^2} \frac{\d{\hat r}}{\d\hat t} + {\dot a} {\hat r},
\label{scaling_v}
\ee
where the dot stands for time differentiation with respect to $t$.
The scaled equation of motion is then
\be
\frac{\d^2 {\hat r}}{\d \hat t^2}+2b^2\left(\frac{\dot a}{a}-\frac{\dot b}{b}\right)\frac{\d {\hat r}}{\d \hat t}+b^4\,\frac{\ddot a}{a}\,{\hat r}=\frac{b^4}{a^3}\, \hat{E}_r \, ,
\label{eqmotion}
\ee
where $\hat E({\hat r},\hat t)$ is the scaled gravitational field. As the density must scale as $\hat\rho({\hat r},\hat t) = a^{3}(t)\rho(r,t)$ in order to preserve the total mass, we scale the gravitational field as $\hat E({\hat r},\hat t)= a^{2}(t)E_r(r,t)$, so that Poisson's equation remains invariant in the scaled variables.

Next, we consider a locally planar perturbation embedded in this expanding universe and denote the corresponding comoving coordinate $\hat x$.
In this locally planar system, Poisson's equation can be approximated by its one-dimensional (1D) counterpart: $\partial_x \hat E = -4\pi G \hat\rho(\hat x, \hat t)$. The simulations presented in the next sections will all be performed in this 1D planar reference frame.

\subsection{Symmetric universe with positive and negative masses}\label{sec:scaling-dm}
In a universe filled with equal amounts of positive-mass and negative-mass matter, the scaling factor $a(t)$ should be linear in time, because over large distances attractive and repulsive gravitational fields cancel each other. Therefore, we take
\be
a(t)= t/t_0,
\label{ACgeneral}
\ee
where $t_0$ is an ``initial" time corresponding to the epoch when the universe becomes transparent to radiation and therefore dominated by matter (in the Dirac-Milne cosmology its value is different from the $3.8\times 10^5$ years after the big bang of standard cosmology \cite{Benoitlevy} -- more details on this will be given in the next section). In the following, the subscript ``0" is used systematically to refer to quantities evaluated at this initial time.
For the function $b(t)$ that determines the scaling of time we use: $b(t)=(t/t_0)^{1/2}$, which has the advantage of rendering the friction term time-independent in Eq. \eqref{eqmotion}. We stress that, while the choice of $a(t)$ is dictated by cosmological considerations, the choice of $b(t)$ is rather arbitrary and can be made with the purely pragmatic aim of making the scaled equations easier to solve, either numerically or analytically.
In this case the relationship between the real time $t$ and the scaled time $\hat t$ is exponential: $t=t_0\exp(\hat t/t_0)$, and the equation of motion becomes
\be
\frac{\d^2 {\hat x}}{\d \hat t^2}+\frac{1}{t_0}\frac{\d {\hat x}}{\d \hat t}= \exp(-\hat t/t_0)\hat{E} .
\label{eqmotionalpha12}
\ee
Note that $\hat t =0$ when $t=t_0$.

Defining the scaled velocity $\hat v$, we obtain Hamilton's equations
\begin{eqnarray}
\frac{\d {\hat x}}{\d \hat t} &=& \hat v \\
\frac{\d \hat v}{\d \hat t} &=& \exp(-\hat t/t_0)\hat{E} - \frac{\hat v}{t_0},
\label{eqmotion_hamilt}
\end{eqnarray}
from which one can deduce the corresponding 1D Vlasov-Poisson system for the scaled distribution functions $F_\pm(\hat r, \hat v, \hat t)$. For the antiplasma and Bondi cases, they read as follows:
\begin{eqnarray}
&&\frac{\partial F_\pm}{\partial \hat t} +
{\hat v}\frac{\partial F_\pm}{\partial {\hat x}} +\frac{m_p^\pm}{m_i^\pm} e^{-\hat t/t_0}\, \hat{E}_\pm\,\frac{\partial F_\pm}{\partial {\hat v}} - \frac{1}{t_0}\frac{\partial ({\hat v}F_\pm)}{\partial {\hat v}}
= 0, \label{vlasov_rescaled} \\
&& \frac{\partial \hat E}{\partial \hat x} = -4\pi G m \,(\hat{n}_{+}-\hat{n}_{-}),
\label{poisson_rescaled}
\end{eqnarray}
where the values of $m_p^\pm$ and $m_i^\pm$ should be read from Table I.

For the Dirac-Milne case, we assume that antimatter (negative mass particles) expands constantly under the repulsive action of all other particles (positive and negative) and can thus be modeled as a uniform repulsive background with $n_{-}(r,t) = n_0/a^3(t)$, where $n_0$ is a constant initial density.
This is not completely exact, as antimatter is also ejected from overdense matter-dominated regions (see Sec. \ref{sec:diracmilne}) leading to a depletion zone, but still constitutes a very good approximation for the study of structure formation.
It is easy to see that, with the scaling defined above, the scaled version of the Vlasov-Poisson equations \eqref{vlasov_diracmilne}-\eqref{poiss_diracmilne1} for the positive-mass particles becomes:
\begin{eqnarray}
&&\frac{\partial F_+}{\partial \hat t} +
{\hat v}\frac{\partial F_+}{\partial {\hat x}} +e^{-\hat t/t_0}\, \hat{E}_+\,\frac{\partial F_+}{\partial {\hat v}} - \frac{1}{t_0}\frac{\partial ({\hat v}F_+)}{\partial {\hat v}}
= 0, \label{vlasov_rescaled_dm} \\
&&\frac{\partial \hat E}{\partial \hat x} = -4\pi G m \,(\hat{n}_{+}-n_0).
\label{poisson_rescaled_dm}
\end{eqnarray}
The forthcoming numerical simulations for the antiplasma, Bondi and Dirac-Milne scenarios, will be based on the solution of Eqs. \eqref{vlasov_rescaled}-\eqref{poisson_rescaled} and \eqref{vlasov_rescaled_dm}-\eqref{poisson_rescaled_dm}.

\subsection{Einstein-de Sitter cosmology}\label{sec:scaling-eds}
For comparison, we briefly review the basic equations for the comoving coordinates in the standard Einstein-de Sitter (EdS) cosmology, without dark energy \cite{Manfredi_PRE2016}.
For a critical universe, the scale factor behaves as: $a(t)=(t/t_0)^{2/3}$. In addition, taking $b(t)=(t/t_0)^{1/2}$, the scaled equation of motion \eqref{eqmotion} becomes autonomous (all coefficients are time-independent):
\be
\frac{\d^2 {\hat x}}{\d \hat t^2}+\frac{1}{3t_0}\frac{\d {\hat x}}{\d \hat t}-\frac{2}{9t_0^2}\,{\hat x}=\hat{E} \, . \label{eqmotion_eds}
\ee
Poisson's equation in $d$ dimensions can be solved exactly for a constant density $n_0$, yielding the gravitational field: $\hat{E}=-\omega_{J0}^2\, \hat x/d$. At steady state, this field must exactly cancel the third term on the left-hand side of Eq. \eqref{eqmotion_eds}, yielding the following relationship between $t_0$ and $\omega_{J0}$:
\be
\omega_{J0}^2 t_0^2 = {2\over 9}d. \label{omgt0-eds}
\ee
The corresponding Vlasov-Poisson equations in comoving coordinates read as \cite{Manfredi_PRE2016}:
\begin{eqnarray}
&&\frac{\partial F}{\partial \hat t} +
{\hat v}\frac{\partial F}{\partial {\hat x}} + \hat{E}\,\frac{\partial F}{\partial {\hat v}} - \frac{1}{3t_0}\frac{\partial ({\hat v}F)}{\partial {\hat v}}
= 0, \label{vlasov_rescaled_RF} \\
&& \frac{\partial \hat E}{\partial \hat x} = -4\pi G m \,(\hat{n}-n_0),
\label{poisson_rescaled_RF}
\end{eqnarray}
where the harmonic term in Eq. \eqref{eqmotion_eds} has been incorporated into Poisson's equation \eqref{poisson_rescaled_RF}.

The fact that all coefficients in Eqs. \eqref{eqmotion_eds} or \eqref{vlasov_rescaled_RF}-\eqref{poisson_rescaled_RF} are time-independent signals that a stationary solution of the scaled equations represents a self-similar solution in the real coordinates.
Note also that, for an expansion in $t^{2/3}$, the product $\omega_J(t)\,t$ is constant in time, i.e.: $\omega_{J}(t) t = \omega_{J0} t_0$.

Another relevant difference between the Einstein-de Sitter and the Dirac-Milne scenarios is that in the latter the scaled gravitational field decreases exponentially with the scaled time, whereas it is constant in the former. An important consequence of this mathematical fact (confirmed by the forthcoming simulations) is that gravitational structure formation in a Dirac-Milne universe stops after a time that is a small multiple of $t_0$, while it continues indefinitely in the Einstein-de Sitter case.

\section{Timescales}\label{sec:timescales}
Before presenting the results of the numerical simulations, it is instructive to identify the principal timescales involved in the problem under study. Two such timescales are the inverse of the initial Jeans frequency  $\omega_{J0}=\sqrt{4\pi G m n_0}$ and the time $t_0$, defined as the time when the universe becomes transparent to radiation (end of recombination epoch). The third timescale is related to the importance of nonlinear effects.

To determine the Jeans frequency, we use the current estimate for today's average baryonic matter density of the universe, i.e. 0.25 protons$/\rm m^{3}$ (there is no need for dark matter in the Dirac-Milne universe):
\[
\rho_{baryon} \rm(now) = 4.18 \times 10^{-28}\, kg/m^3.
\]
Following Benoit-L{\'evy} and Chardin \cite{Benoitlevy}, recombination should occur at $t_0=14 \times 10^6$ years in a Dirac-Milne universe, compared to $380\,000$ years for standard cosmology. Thus, $t_{now} = 14\times 10^9 \,{\rm y} \approx 10^3 t_0$, and since the expansion factor is linear in $t$, today's mass density should be multiplied by $10^9$ in order to obtain $\rho_0$ at $t=t_0$. Finally we get:
\be
\omega_{J0}\, t_0 \approx 8.3. \label{omgt0}
\ee
At the present epoch, this parameter takes the value: $\omega_{J, now}\, t_{now} \approx 0.26$.

A third timescale is given by the initial density fluctuations, which determines the typical timescale over which nonlinear effects become important (known as the bounce time in plasma physics \cite{Ryutov99}). This nonlinear time $t_{nl}$ can be written as:
\be
\omega_{J0} t_{nl} = \sqrt{n_0/\tilde{n}_\lambda}
\ee
where $\tilde{n}_\lambda$ is the amplitude of the initial density fluctuations (which depends on their wavelength $\lambda$).

In summary, we have identified three distinctive timescales for this problem:
\begin{itemize}
\item
The inverse of the Jeans frequency $\omega_{J0}$, which determines the rate of the Jeans instability, see Eqs. \eqref{omega_anti} and \eqref{omega_dm};
\item
The recombination time $t_0$, which acts as an initial time in our model. Note that if $\omega_{J0}\,t_0 \ll 1$ the friction term in Eq. \eqref{eqmotionalpha12} dominates from the start, so that cosmological structures will never have time to form. In contrast, if $\omega_{J0}\,t_0 \gg 1 $, structures will keep forming for very long times. The value $\omega_{J0}\, t_0 \approx 8.3$ that we find for the Dirac-Milne universe is sufficiently large to allow for structure formation at the beginning of the matter-dominated epoch. Such structure formation should stop at a time ($\approx 10^9\, \rm y$) relatively close to the present epoch, as will be shown in the forthcoming simulations;
\item
The nonlinear timescale $t_{nl}$, which plays an important role in the Bondi scenario, for which no Jeans-like instability exists, and gravitational structures can only form via nonlinear couplings. Note that this ``parameter" is actually a dynamical quantity, as the level of fluctuations changes in time in the case of an instability. Thus, its initial value may not be the relevant one for the later evolution in the antiplasma and Dirac-Milne scenarios, where the initial stages of the evolution are dominated by the Jeans instability.
\end{itemize}
In the forthcoming numerical simulations, we will use units in which $n_0=\omega_{J0}=1$, so that the relevant dimensionless parameters are $\omega_{J0} t_{0}$ and $\omega_{J0} t_{nl}$.

Note that in the EdS scenario there are only {\em two} relevant timescales, because the product $\omega_{J0}t_0=\sqrt{2/3}\approx 0.82$ is fixed by Eq. \eqref{omgt0-eds}. Further, in the EdS case the product $\omega_{J}(t)\,t = \omega_{J0}t_0$ is an invariant, i.e. it is the same for all epochs, whereas it decreases (as $t^{-1/2}$) in the Dirac-Milne case.
This simple fact illustrates a fundamental difference between the EdS and Dirac-Milne universes. In the former, $\omega_{J}(t)\,t$ is of order unit at all epochs, so that structure formation occurs all along the life of this type of universe. In contrast, the Dirac-Milne universe starts with a relatively large value $\omega_{J0}t_0=8.3$, which then decreases and becomes lower than unity near the present epoch, so that structure formation eventually stops.

To put it somewhat figuratively: the EdS universe is a universe that never ages and produces ``offsprings" (i.e., structures) during its entire lifetime. The Dirac-Milne universe ages with time and after a certain epoch (corresponding roughly to $\omega_{J}(t)\,t \approx 1$) it becomes ``sterile" and begets no further gravitational structures.

\section{Numerical results}\label{sec:numres}
In this section we present the results of numerical simulations of the three models with negative mass considered so far, namely the antiplasma, Bondi, and Dirac-Milne scenarios, with particular emphasis on the latter case. For comparison, we will also show some results obtained with a standard Einstein-de Sitter cosmology. The simulations were performed using an N-body code that solves the equations of motion \eqref{eqmotionalpha12} for $N$ interacting particles. For $N\to \infty$ the N-body problem tends to its mean-field limit, described by the Vlasov-Poisson equations discussed in the preceding sections. Typical simulations employed $N =2.5\times 10^5$ particles.

As mentioned in the preceding sections, we consider a 3D expanding spherically-symmetric universe and then study planar perturbations in the comoving coordinates. This effectively reduces the problem to one spatial dimension in the local comoving coordinate $\hat x$, which will be represented in the numerical results. More details on the model can be found in Refs. \cite{MRexp,MR2010,miller2010ewald,Manfredi_PRE2016}.
Since we use a 1D approximation, the ``particles" are in fact infinite sheets of mass $m$ (in absolute value). Boundary conditions are taken to be spatially periodic with period $L$. The initial velocity spread is very small (except for the Bondi case which, lacking a linear instability, must be strongly excited in order to observe some interesting dynamics).

In all presented results, time is expressed in years elapsed since the Big Bang ($t=0$). Density is expressed in terms of $n_0$. The units of space and gravitational field are somewhat arbitrary: once we have fixed a certain unit of time (say $\omega_{J0}^{-1}$, as was done in the code), space is measured in units of a certain length $\lambda$ and the gravitational field in units of $\lambda\omega_{J0}^{2}$. Thus, $\lambda$ disappears from the scaled equations of motion \eqref{eqmotion} and \eqref{eqmotion_eds}: one chooses an arbitrary value for it, and then all lengths are measured in terms of this unit.

In all cases, the initial density is the sum of a spatially uniform term $\rho_0=mn_0$ and a small perturbation $\tilde\rho$ with power spectrum $|\tilde\rho_k|^2 \sim k^p$, where $k$ is the wavenumber. Initial power spectra of this form, with $p \in [0,4]$, were used in a number of earlier works on structure formation \cite{Joyce2011,MR2010}.
In the present work, we take $p=2$, which produces a spectrum that is largest at small wavelengths, and then study the clustering of matter at increasingly larger scales, until the size $L$ of the box is reached.

We start by briefly showing a few results on the antiplasma and Bondi cases, then turn to a more extensive comparison between the Dirac-Milne and Einstein-de Sitter cosmologies.

\subsection{Antiplasma and Bondi cases}\label{sec:numres-antibondi}
As was discussed in Sec. \ref{sec:linear-anti}, the antiplasma case is subject to a Jeans instability with growth rate of the order of the Jeans frequency $\omega_{J0}$. Consequently, the system quickly departs from the initial equilibrium and develops nonlinear structures. These structures are shown in snapshots of  the matter density and  phase-space distributions taken at different instants, as depicted in Fig. \ref{fig:AP-phasespace}.

\begin{figure}[]
 \begin{center}
\includegraphics[width=\linewidth]{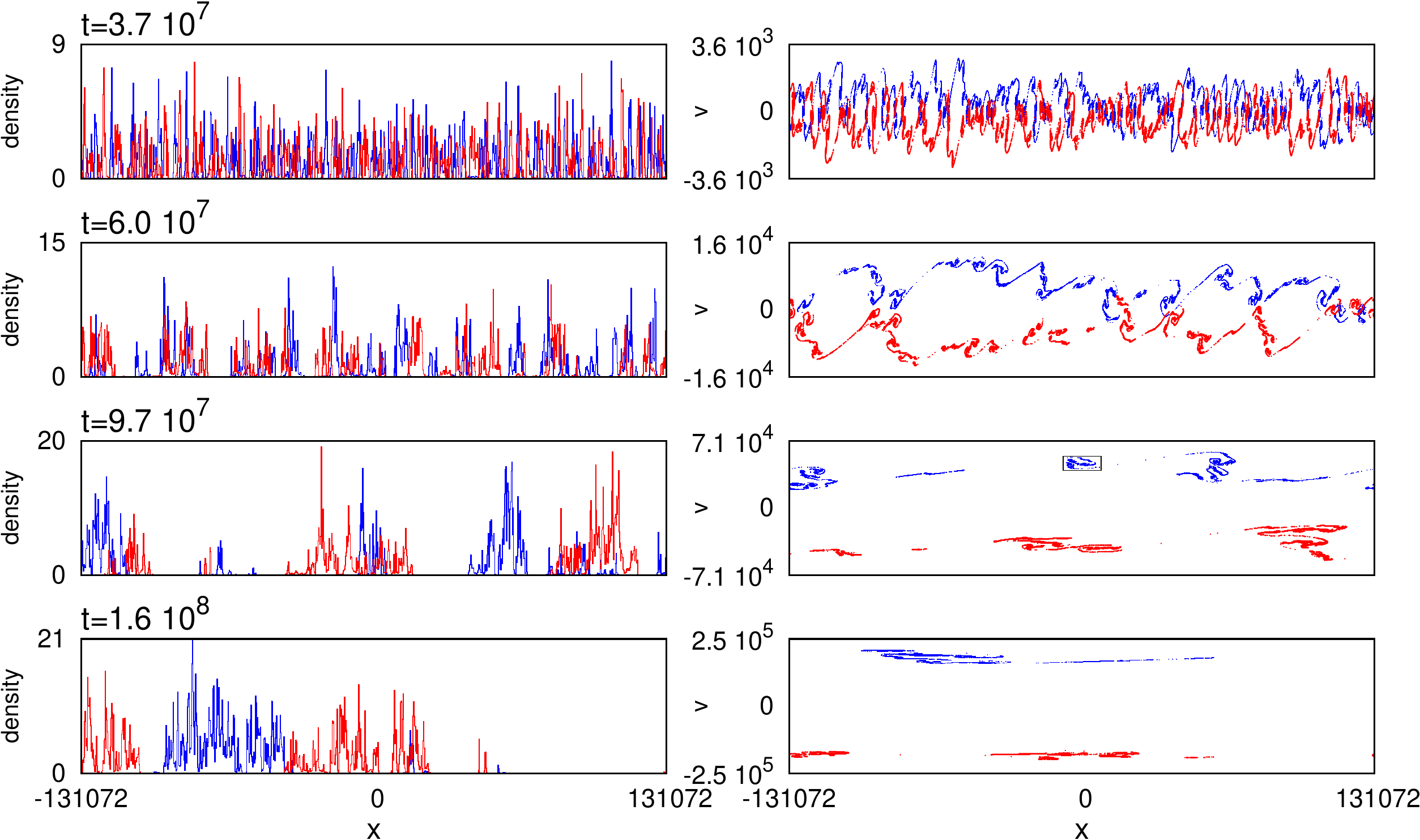}
 \end{center}
    \caption{Antiplasma case. Left frames: Matter density $\rho(x,t)$ for positive mass (red) and negative mass (blue) particles at four different times. Right frames: Corresponding distributions in the scaled phase space. The small rectangle (right column, third panel) indicates the region that is zoomed in on Fig. \ref{fig:AP-zoom}. Time is measured in years elapsed after the Big Bang.} \label{fig:AP-phasespace}
\end{figure}

As particles of like masses attract and particles of unlike masses repel, we observe the segregation of clumps of positive and negative matter, visible in the density plots.
Interestingly, there is also a segregation in velocity space, with two counterpropagating streams made of positive and negative particles appearing. Thus, the Jeans instability extracts gravitational potential energy from the system (initially at rest in the comoving frame) and converts it into kinetic drift energy. Nevertheless, by zooming in on one of the streams, one still observes plenty of intricate gravitational structures, as is shown in Fig. \ref{fig:AP-zoom}.
Here and in other forthcoming figures, the zoom region changes with time because it follows the trajectories of the particles that are contained initially in a given region of the phase space at a chosen time (here, the small rectangle in the phase space plot in Fig. \ref{fig:AP-phasespace}.

\begin{figure}[]
 \begin{center}
\includegraphics[width=\linewidth]{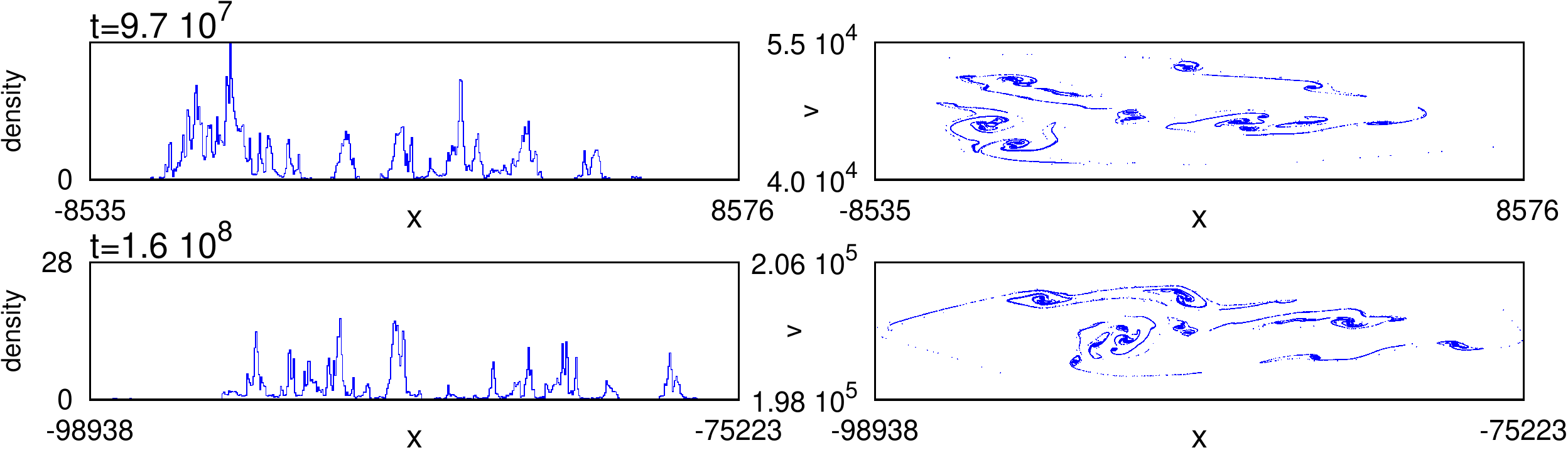}
 \end{center}
    \caption{Antiplasma case. Zooms of the region denoted by a small rectangle in Fig. \ref{fig:AP-phasespace} (right column, third panel).
    Left frames: matter density; Right frames: phase space.} \label{fig:AP-zoom}
\end{figure}

This behavior may seem counterintuitive, but performing the linear analysis of an equilibrium consisting of two counterpropagating streams, reveals that such equilibrium is (for large enough drift velocities) stable, in contrast to the one-stream initial conditions which is unstable as was shown in Sec. \ref{sec:linear-anti}. Thus, the system evolves from an unstable equilibrium (one stream with zero mean velocity) to a stable equilibrium (two streams with finite and opposite velocities). Note also that this is exactly the opposite situation compared to a two-component plasma, for which one-stream equilibria are stable whereas two-stream equilibria are unstable \cite{Chen,ManfrediRouet}.
This peculiar behavior in the phase space constitutes one further reason to rule our the antiplasma scenario for cosmological applications.

\begin{figure}[]
 \begin{center}
\includegraphics[width=\linewidth]{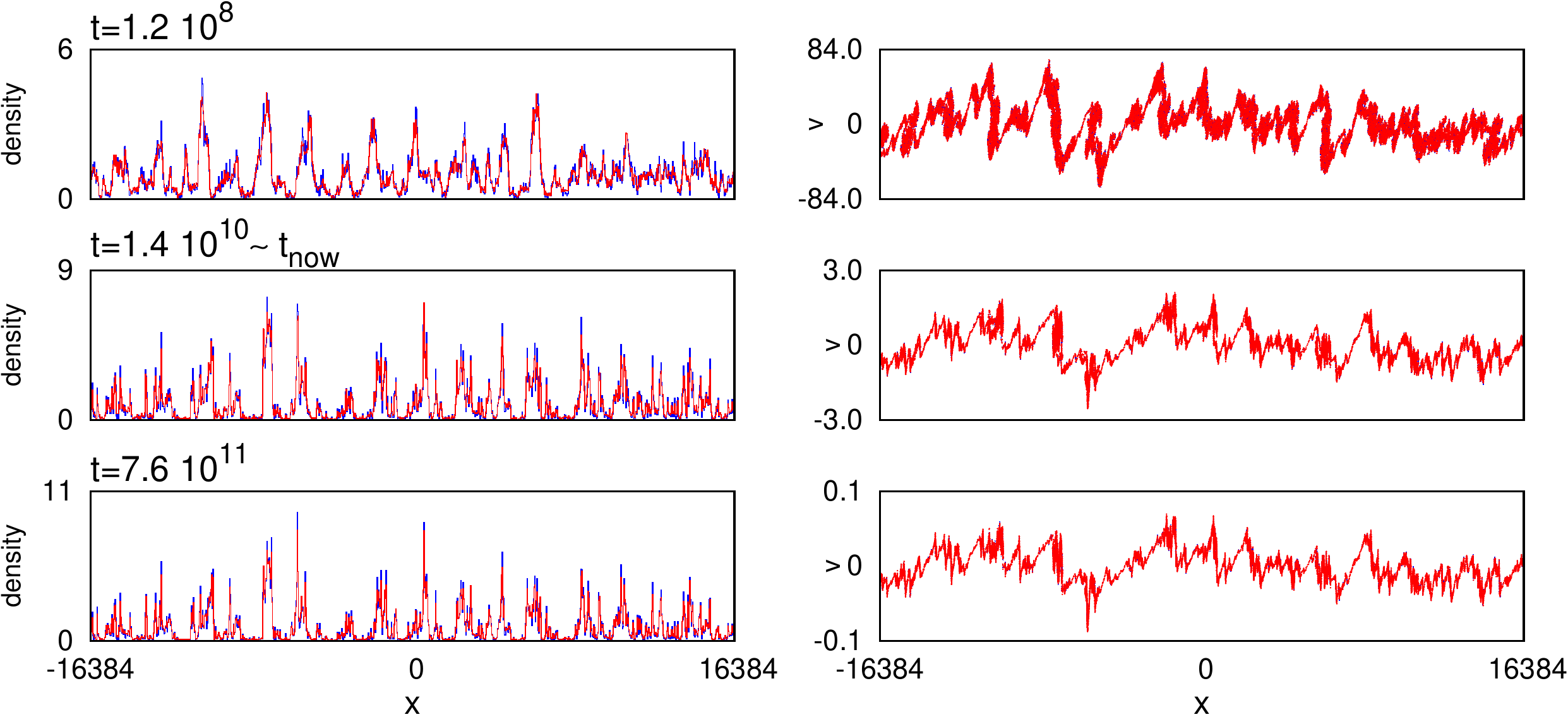}
 \end{center}
    \caption{Bondi case. Left frames: Matter density $\rho(x,t)$ for positive mass (red) and negative mass (blue) particles at three different times. Right frames: Corresponding distributions in the scaled phase space. Time is measured in years elapsed after the Big Bang.} \label{fig:Bondi-phasespace-8}
\end{figure}

For the Bondi case, as we saw in Sec. \ref{sec:linear-bondi}, there is no linear instability. Therefore, we need to introduce very large fluctuations in the system's initial condition in order to observe some structure formation. If the initial fluctuations are low, then no gravitational structures are formed in the subsequent evolution.
The most salient feature observed in the simulations is that, in contrast to the antiplasma case, positive and negative particles clump together to form ``neutral" regions (see Fig. \ref{fig:Bondi-phasespace-8}, which only shows one eighth of the total computational box for clarity). Also the structures form very early in the evolution and remain basically unchanged thereafter.

\subsection{Dirac-Milne and Einstein-de Sitter universes} \label{sec:numres-dm}

We now turn to the more interesting case of the Dirac-Milne universe and its comparison with the Einstein-de Sitter case. As was mentioned in Sec. \ref{sec:scaling-dm}, one can simplify the treatment by assuming that the negative-mass component remains uniform in the comoving coordinates, so that only positive-mass particles need to be simulated. The initial conditions are identical for both sets of simulations, although the physical initial time ($t_0$, representing the recombination epoch) is of course not the same.

\begin{figure}[]
 \begin{center}
\includegraphics[width=\linewidth]{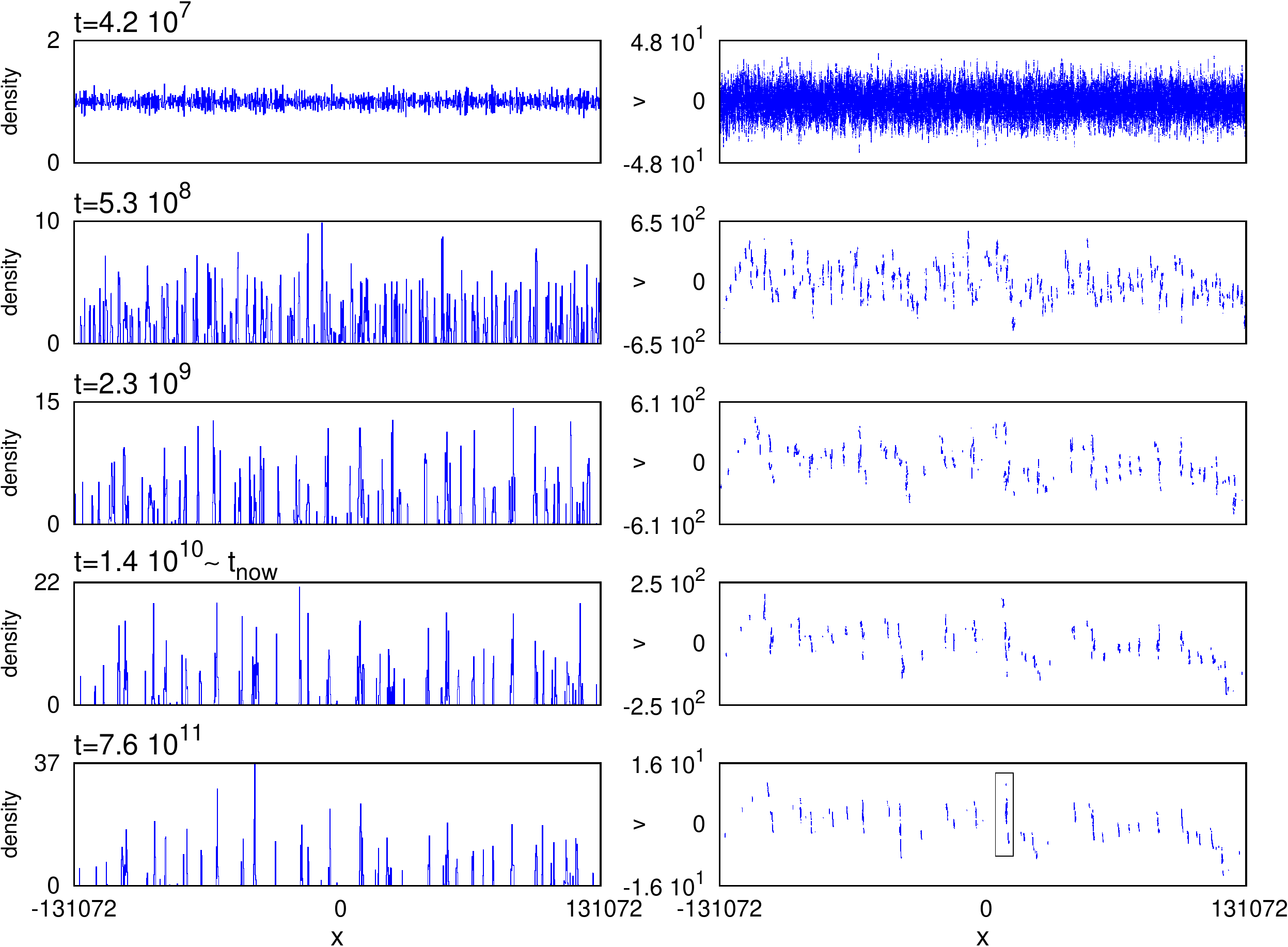}
 \end{center}
    \caption{Dirac-Milne case. Left frames: Matter density $\rho(x,t)$ for the positive-mass particles at five different times. Right frames: Corresponding distributions in the scaled phase-space. The fourth set of frames from the top corresponds roughly to the present epoch.} \label{fig:DiMi-phasespace}
\end{figure}

\begin{figure}[]
 \begin{center}
\includegraphics[width=\linewidth]{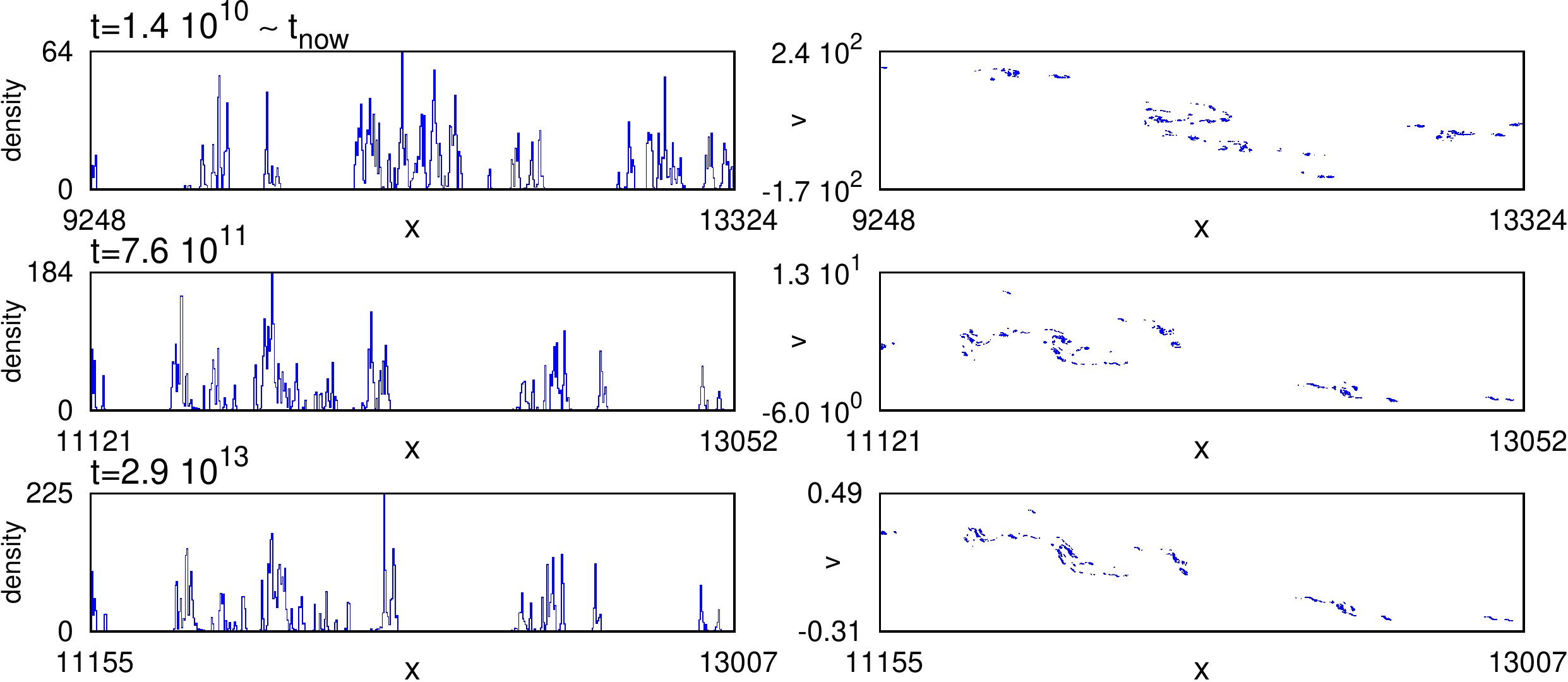}
 \end{center}
    \caption{Dirac-Milne case. Left frames: Matter density $\rho(x,t)$ for the positive-mass particles at three different times. Right frames: Corresponding distributions in the scaled phase-space. The figures are zooms of the region denoted by a small rectangle in Fig. \ref{fig:DiMi-phasespace} (bottom right frame).} \label{fig:DiMi-zoom}
\end{figure}

The matter density and phase-space distributions for the Dirac-Milne case are provided in Fig. \ref{fig:DiMi-phasespace} and a zoom is shown in Fig. \ref{fig:DiMi-zoom}. At first sight, structure formation occurs similarly to the standard (EdS) case, driven by the Jeans instability \cite{MRexp,MR2010,miller2010ewald,Manfredi_PRE2016}.
However, some crucial differences are noticeable.
{
In particular, structure formation starts earlier and stops before the present epoch. Indeed, almost no significant changes can be seen in the density and phase-space distributions from the present epoch onwards.
}
Mathematically, this is due to the fact that, in the comoving equations of motion \eqref{eqmotion_hamilt}, the relative impact of the friction term increases with respect to the gravitational field. Indeed, the scaled velocity in the phase-space plots of Fig. \ref{fig:DiMi-phasespace} initially increases under the action of the Jeans instability, but later decreases rapidly because of the effect of the friction term in the comoving equations of motion. Structure formation halts when this latter term becomes dominant.
\begin{figure}[]
 \begin{center}
\includegraphics[width=\linewidth]{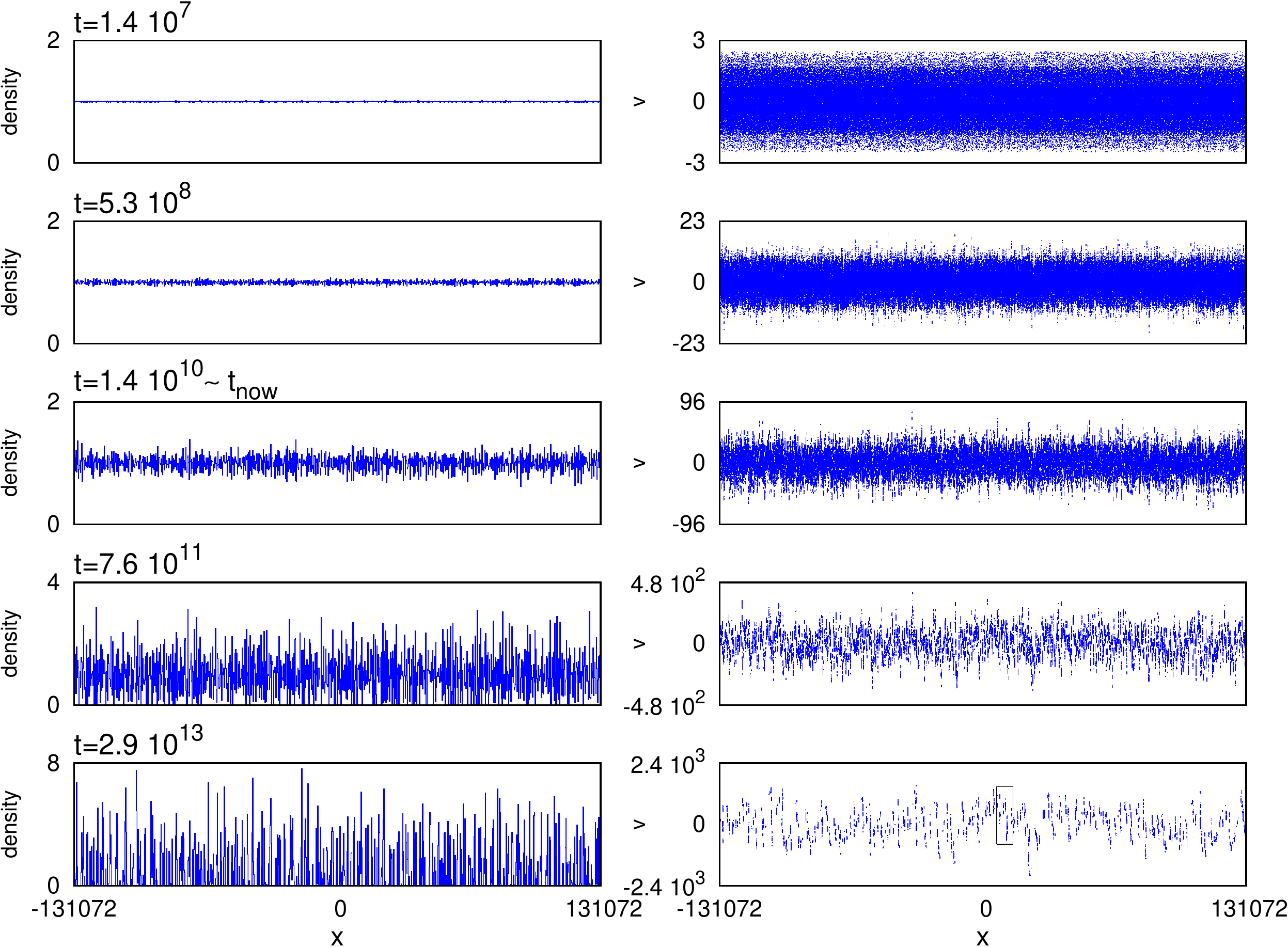}
 \end{center}
    \caption{Einstein-de Sitter case. Left frames: Matter density $\rho(x,t)$ for the positive-mass particles at five different times. Right frames: Corresponding distributions in the scaled phase-space. The third set of frames from the top corresponds roughly to the present epoch.} \label{fig:EdS-phasespace}
\end{figure}

\begin{figure}[]
\includegraphics[width=\linewidth]{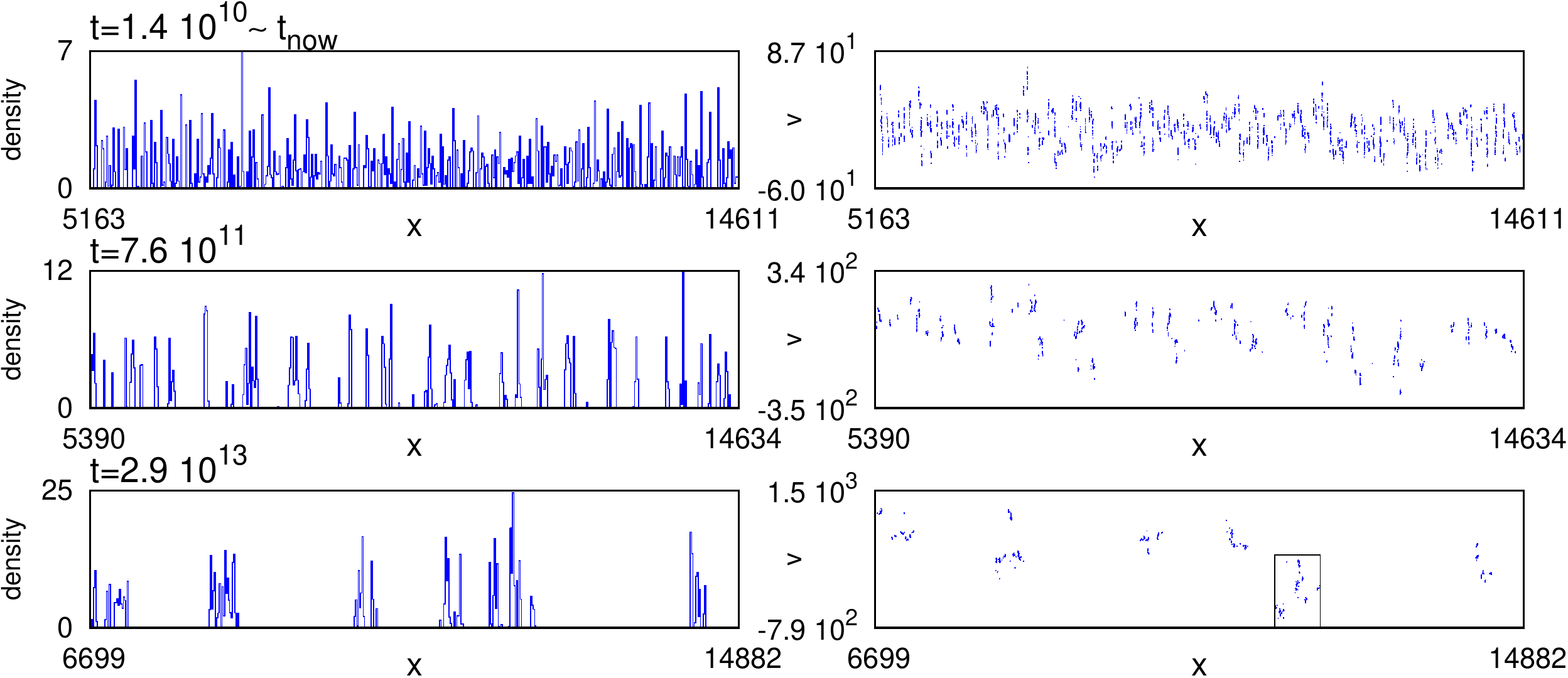}
    \caption{Einstein-de Sitter case. Zooms of the region denoted by a small rectangle in Fig. \ref{fig:EdS-phasespace} (bottom right frame).}
    \label{fig:EdS-zoom1}
\end{figure}

\begin{figure}[]
\includegraphics[width=\linewidth]{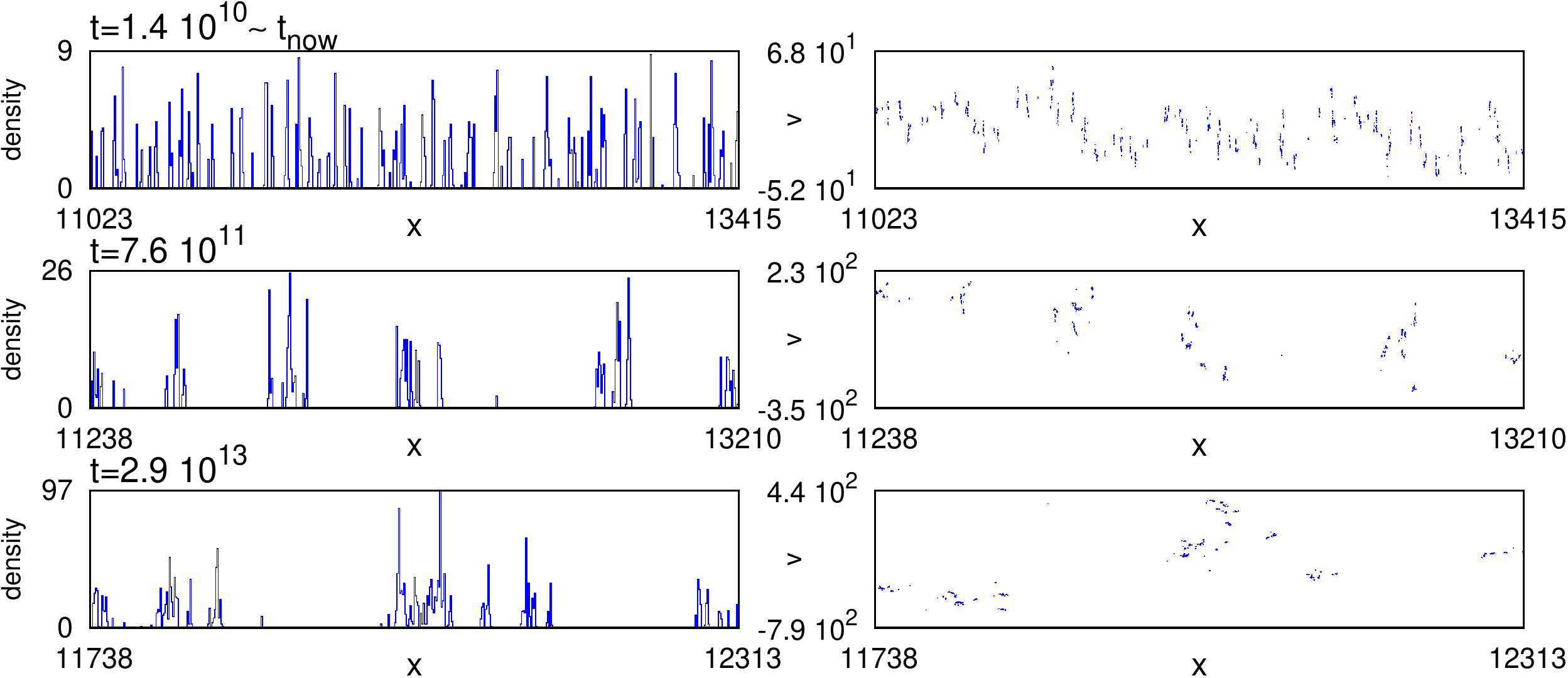}
    \caption{Einstein-de Sitter case. Further zooms of the region indicated by a rectangle on Fig. \ref{fig:EdS-zoom1}.} \label{fig:EdS-zoom2}
\end{figure}

Let us now compare this behavior with that observed for the Einstein-de Sitter universe, which was already simulated in earlier works \cite{MRexp,MR2010,miller2010ewald,Manfredi_PRE2016}.
The matter density and phase-space distributions are plotted in Fig. \ref{fig:EdS-phasespace}, and two consecutive zooms in Figs. \ref{fig:EdS-zoom1} and \ref{fig:EdS-zoom2}. In accordance with previous results, the formation of gravitational structures continues for all times, as is particularly clear from the two zooms. The self-similar nature of the matter distribution in the phase space is also apparent: two snapshots taken at different scales (but at the same time) look qualitatively similar, indicating that there is no intrinsic spatial scale for this system. This fact has led us, in earlier works, to compute a fractal dimension for the matter distribution \cite{MR2010,Manfredi_PRE2016}. These features are the hallmark of a critical universe.

This different behavior is in accordance with our discussion of Sec. \ref{sec:timescales}, where we pointed out that the Dirac-Milne universe ``ages" with time, meaning that it ceases to produce gravitational structures. In contrast, the EdS universe does not age and continuously generates new structures at all scales.

\begin{figure}[]
\includegraphics[width=0.6\linewidth]{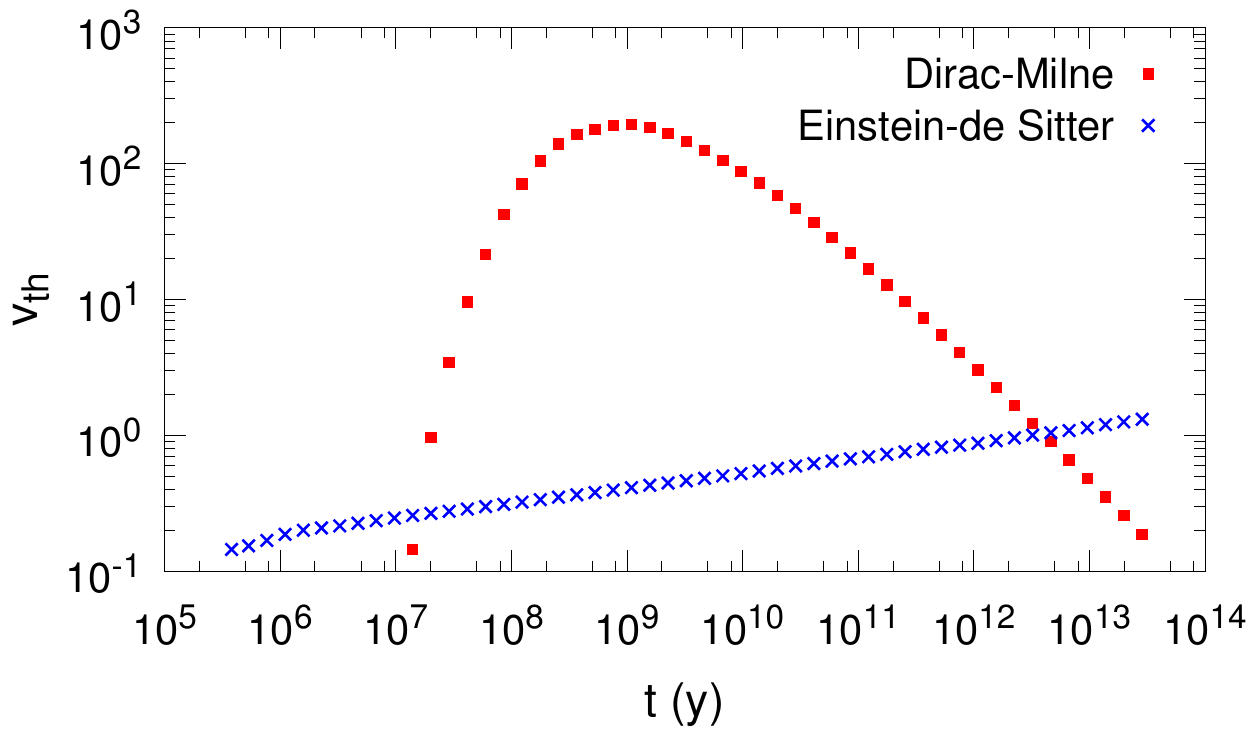}
    \caption{Time evolution of the thermal peculiar velocities, for the EdS and Dirac-Milne universes. } \label{fig:vth}
\end{figure}

Another difference between the Dirac-Milne and EdS cases is that in the former the (scaled) velocity range of the particles decreases after an initial growth  (Fig. \ref{fig:DiMi-phasespace}), whereas in the latter it constantly increases (Fig. \ref{fig:EdS-phasespace}). This signals a change in behavior for the Dirac-Milne universe, which should roughly correspond to the epoch where structure formation ends. In order to understand the relevance of this quantity, let us go back to the definition of the scaled velocity, Eq. \eqref{scaling_v}, which can be rewritten as
\be
v = \frac{a}{b^2} {\hat v} + H(t) r ,
\label{scaling_v-dm}
\ee
where $H(t)=\dot a/a$ is the Hubble parameter.
The first term on the right-hand side of Eq. \eqref{scaling_v-dm} represents the so-called peculiar velocity of the masses (residual velocity in the comoving reference frame), while the second is the Hubble redshift term.
In Fig. \ref{fig:vth}, we plot the evolution of the ``thermal" peculiar velocity, defined as $v_{th} \equiv \langle (v-Hr)^2\rangle^{1/2}$, where the average is taken over all the particles. The difference is clear: for Dirac-Milne, the spread of the peculiar velocities first increases, reaches a peak, and then decreases continuously. Structure formation halts when $v_{th}$ starts decreasing, i.e., around $t_c \approx 10^9$ years.
We also note that this time corresponds roughly to the time when the parameter $\omega_J(t)\,t$ reaches unity. Indeed, for the Dirac-Milne case:
\be
\omega_J(t)\,t = \omega_{J0}\,t_0 \, \left({t \over t_0}\right)^{-1/2} \approx 8.3\, \left({t \over t_0}\right)^{-1/2}
\label{eq:omjt}
\ee
becomes equal to unity for $t_c \approx 69 t_0 \approx 10^9\, \rm y$. This is in agreement with our earlier discussion in Sec. \ref{sec:timescales}.

Peculiar velocities of clusters of galaxies have been estimated by measuring their impact on the cosmic microwave background spectrum \cite{Kashlinsky2008}. A compilation of several peculiar velocity surveys was published in recent years \cite{Watkins2009}. These surveys consistently showed that the observed velocities exceed the prediction of the $\Lambda\rm CDM$ model by a factor of $\approx 5$. Although the discrepancy between the Dirac-Milne and EdS predictions is even higher, our result (Fig. \ref{fig:vth}) goes in the right direction. A proper comparison between the $\Lambda\rm CDM$ and Dirac-Milne cosmologies will be studied in a forthcoming work.

\begin{figure}[]
\includegraphics[width=0.7\linewidth]{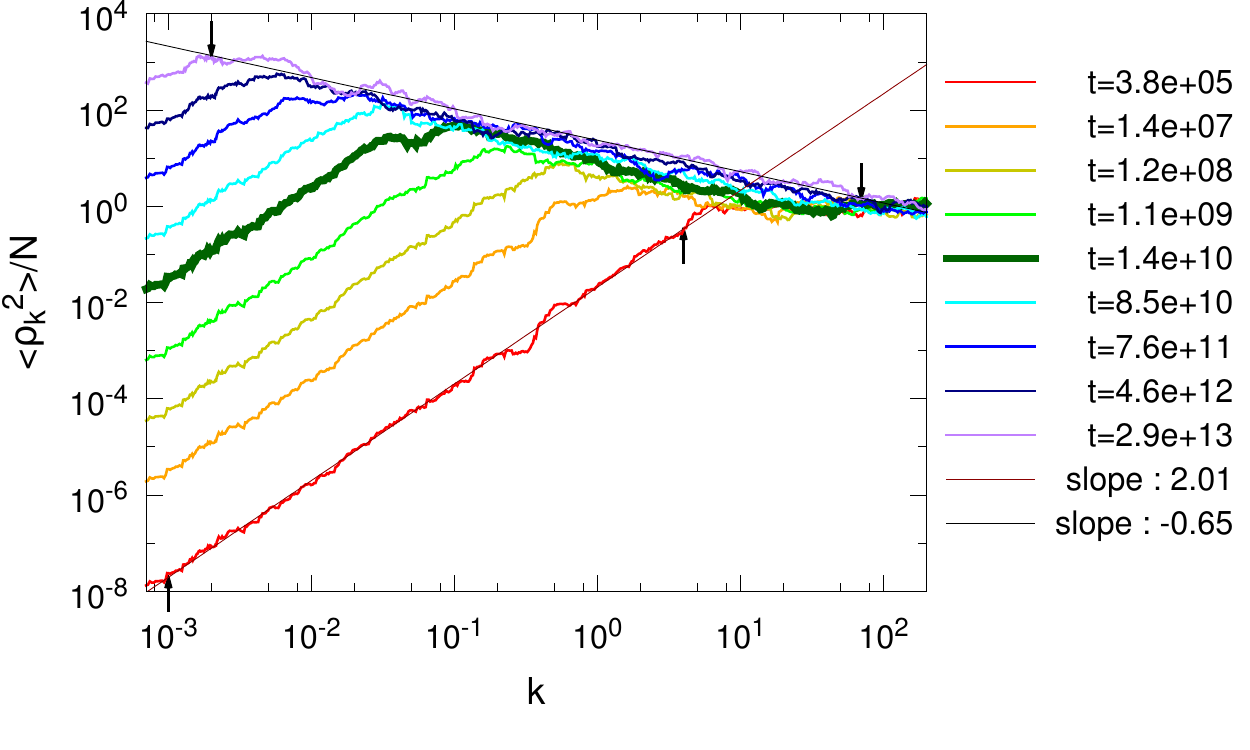}
\includegraphics[width=0.7\linewidth]{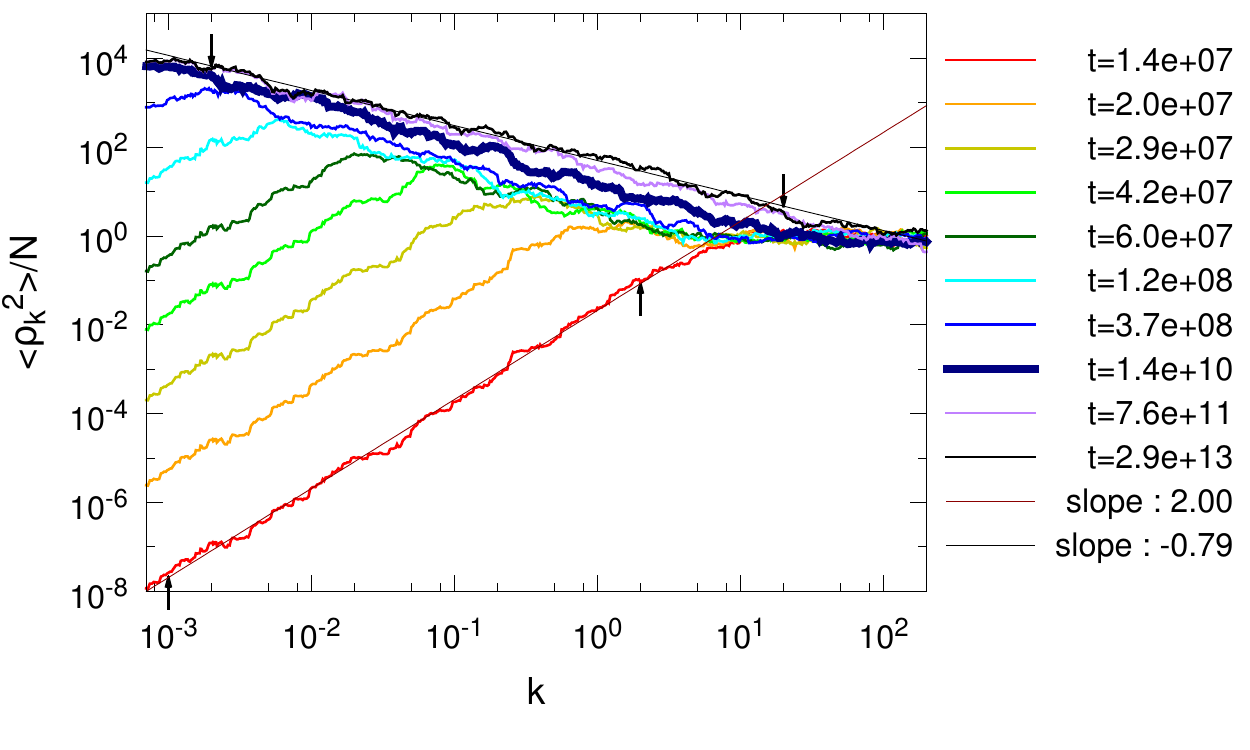}
    \caption{Power spectra of the matter density for the EdS (top panel) and Dirac-Milne (bottom panel) universes. Time is expressed in years after the Big Bang. The thick lines correspond to the present epoch ($t= 14\times 10^9$y).} \label{fig:spectra}
\end{figure}

\begin{figure}[]
\includegraphics[width=0.7\linewidth]{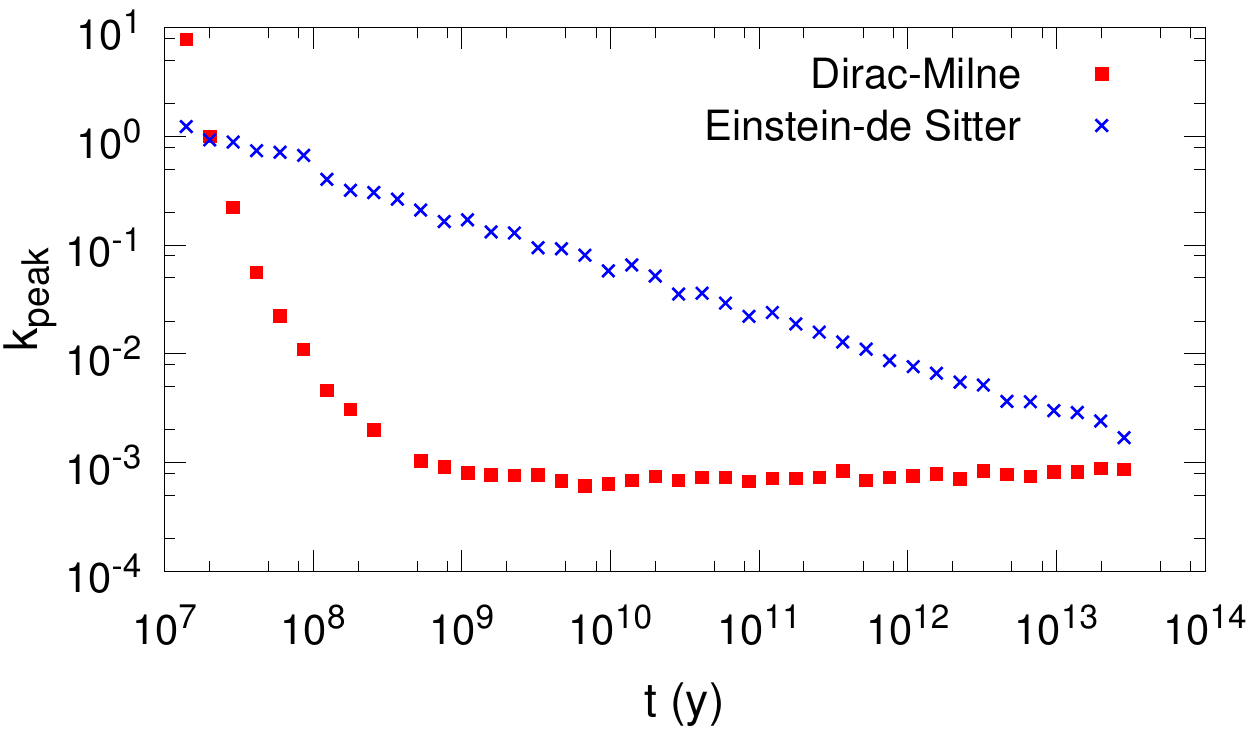}
    \caption{Wavenumber corresponding to the peak in the power spectrum for the Dirac-Milne and Einstein-de Sitter universes as a function of time, in comoving coordinates.} \label{fig:kmin}
\end{figure}

Finally, we show in Fig. \ref{fig:spectra} the power spectra of the matter density for the EdS  and Dirac-Milne cases. Both spectra begin (at $t=t_0$) as a power law $|\tilde\rho_k|^2 \sim k^p$ with $p= 2$ up to $k_{max}= 2\pi/d$, where $d\approx 1$ is the average initial interparticle distance. For larger wavenumbers the spectrum is flat (white noise), simply because there are not enough particles to resolve such short distances.

{\modif
Initially, the low wavenumber amplitudes (leftmost part on Fig. \ref{fig:spectra}) are amplified but keep their shape intact: this is the linear stage of the instability.
The linear growth rate of gravitational structures observed in these simulations is significantly larger for the Dirac-Milne case compared to the EdS one. However, it is also well-known that the linear growth is faster for the concordant $\Lambda\rm CDM$ model than for the EdS case -- see e.g. Ref. \cite{Huterer2015} -- so that our Dirac-Milne result again appears to go, at least qualitatively, in the right direction.
}

As time increases, nonlinearities start to play a predominant role and the spectra take a characteristic shape that is qualitatively similar both in the Dirac-Milne and the EdS cases, as can be seen from Fig. \ref{fig:spectra}. At low wavenumbers (region I), the spectrum is still of the type $|\tilde\rho_k|^2 \sim k^2$, a remnant of the initial condition. At larger wavenumbers, an intermediate region (II) appears, characterized by a power-law spectrum with negative exponent, slightly steeper for the Dirac-Milne ($p=-0.78$) compared to the EdS ($p=-0.67$) case. This power-law region is rather robust (it extends over roughly four decades) and constitutes the signature of hierarchical clustering in the phase space, as was seen in the previous plots (Figs. \ref{fig:DiMi-phasespace} and \ref{fig:EdS-phasespace}). For even larger wavenumbers, the spectra are again flat (region III).

The spectra display a peak separating regions I and II, which can be taken as the typical cluster size at a certain epoch. In both cases, the peak initially moves towards smaller and smaller wavenumbers (larger scales), although quite noticeably faster for the Dirac-Milne case (Fig. \ref{fig:kmin}). But for longer times ($> 10^9\, \rm y$) the behaviors diverge: for EdS, the position of the peak $k_{peak}(t)$ keeps moving to larger and larger scales (eventually reaching the size of the computational box, at which point the simulation would not be valid anymore), whereas for Dirac-Milne it saturates at a constant value. A similar saturation effect is also expected for $\Lambda \rm CDM$.
At the present epoch, the typical cluster size in comoving coordinates ($\ell \sim k_{peak}^{-1}$) is almost two orders of magnitude larger for the Dirac-Milne universe compared to the EdS case. We also stress that such cluster size is determined primarily by nonlinear effects occurring during the matter-dominated epochs.

In summary, Figs. \ref{fig:spectra}--\ref{fig:kmin} clearly show that, compared to EdS, structure formation begins earlier and proceeds initially faster in the Dirac-Milne scenario. However, the production of new gravitational structures does not continue forever (as in EdS), but rather stops a few billion years after the Big Bang.
This confirms the intuition that structure formation is linked to the dimensionless parameter $\omega_J(t)t$, as mentioned in Sec. \ref{sec:timescales}. For the Dirac-Milne case, this quantity is large at recombination ($\omega_{J0}t_0 \approx 8.3$) so that structures initially form quickly and efficiently; then it decreases (as $t^{-1/2}$) and, when it falls below unity, structure formation halts.
In contrast, for EdS, $\omega_J(t)t =\omega_{J0}t_0 = \sqrt{2/3}$ is constant and of order unity for all times, so that structure formation occurs at a moderate rate for all epochs.

As a closing remark to this section, we note that here we compared the Dirac-Milne universe to the EdS case mainly for reasons of computational simplicity. This should be accurate enough for matter-dominated epochs (i.e., between recombination and the vacuum epoch), after which the full $\Lambda\rm CDM$ model should be used instead of the simpler EdS.
Both the Dirac-Milne and the $\Lambda\rm CDM$ models describe a universe that ``ages" with time, meaning that structure formation halts after a certain time (see Sec. \ref{sec:timescales}). Interestingly, the Dirac-Milne cosmology can reproduce this feature of the observed universe without resorting to dark energy. The timescales of structure formation in the Dirac-Milne and $\Lambda\rm CDM$ universes are not too different either. Indeed, for $\Lambda\rm CDM$, structure formation ends when:
\(
a(t)/a(t_{now}) = \left(t/t_{now}\right)^{2/3} = \left(\Omega_m/\Omega_\Lambda\right)^{1/3},
\)
which (using the current estimations for the normalized densities: $\Omega_m=0.3$, $\Omega_\Lambda=0.7$) yields $t = t_{\Lambda} \approx 0.65\,t_{now} \approx 9\times 10^9\,\rm y$.
This is larger than the equivalent ``freezing" time $t_c \approx 10^9\, \rm y$ defined above for the Dirac-Milne cosmology, but in both cases structure formation ended before the present epoch. Also keep in mind that this is a rather crude estimate of $t_c$. A closer inspection of the phase-space portraits (Fig. \ref{fig:DiMi-phasespace}, and other snapshots not shown here) suggests that structure formation stops some time between $10^9$ and $10^{10}$ years after the Big Bang, which is approximately  compatible with $\Lambda\rm CDM$.

\section{Discussion and conclusions}\label{sec:conclusion}

The scope of this work was twofold. Firstly , we developed a general formalism that allows for the existence of negative masses in a Newtonian framework. The family of models that we obtained includes not only those that can be defined by the different signs of the active, passive, and inertial masses, but also a more general class that can only be characterized by a set of {\em two} Poisson's equations for the gravitational potential. These models could be viewed as the Newtonian limit of some bimetric extension of General Relativity.
There is a total of seven nontrivial models that can be constructed in this way.

We produced circumstantial evidence that only one of these models possesses the right features that make it compatible with the alternative ``Dirac-Milne" cosmology recently proposed by  Benoit-Levy and Chardin \cite{Benoitlevy}. This scenario assumes a universe composed of equal amounts of matter and antimatter, in which the latter antigravitates. In the model that we propose to simulate the Dirac-Milne scenario, all gravitational interactions between matter and antimatter are repulsive, except for matter-matter interactions, which must of course be attractive.
These features imply that antimatter is ejected from overdense matter-dominated regions (galaxies), while spreading out in underdense regions to form a uniform dilute repulsive background, with very few annihilation events occurring, in accordance with observations.

In a cosmological context, it is this repulsive background that induces an expansion rate linear in time, $a(t) \sim t$, for the Dirac-Milne universe, faster than the expansion rate of the matter-dominated epochs in the standard cosmological model, which scales as $a(t) \sim t^{2/3}$. Hence, the Dirac-Milne universe does not suffer from the horizon problem and does not need primordial inflation to explain the current homogeneity at large scales. Indeed, the dilute repulsive background could be viewed as a cosmological constant that decreases in time, with the corresponding vacuum energy decreasing volumetrically as $a^{-3}$, see Eqs. \eqref{vlasov_rescaled_dm}-\eqref{poisson_rescaled_dm}.

The second purpose of this work was to study the implications of the Dirac-Milne cosmology on gravitational structure formation. To do this, we devised a local 1D model embedded in a 3D spherically expanding universe and performed N-body simulations of both a Dirac-Milne and an Einstein-de Sitter ($\Omega_m=1, \, \Omega_\Lambda=0$) universe.
In both cases we observed gravitational structure formation, with clusters and subclusters developing from an almost uniform initial condition. After the nonlinear stage of the evolution is reached, both models display a power-law behavior in the wavenumber spectrum of the matter density -- with, however, one crucial difference: whereas for EdS the formation of structures continues indefinitely, for the Dirac-Milne universe it stops a few billion years after the Big Bang, after which time the gravitational structures stay frozen in comoving coordinates. Today, the typical size of the structures should be over an order of magnitude larger in the Dirac-Milne universe.

In this work, for the sake of computational simplicity, we compared the Dirac-Milne universe to the Einstein-de Sitter case, which should be accurate for matter-dominated epochs  before the effect of the cosmological constant has become dominant.
Nevertheless, for a proper analysis of structure formation, the Dirac-Milne scenario should be compared to the standard $\Lambda \rm CDM$ model. It should be possible to do this in the framework of the present 1D approach, thus keeping the computational cost relatively low. The results of such comparison will be reported in future works.
For the time being, we can conclude that the Dirac-Milne universe generates hierarchical structures compatible with those observed in our universe, and whose formation halts at an epoch earlier than the present one, as is also the case for the standard cosmological model.
All in all, these preliminary results are encouraging and deserve to be further verified and quantitatively compared both with observations and with the predictions of the standard $\Lambda \rm CDM$ scenario.

\noindent{\bf \large Acknowledgments}\\
The numerical simulations were partly performed on the computer cluster at the Centre de Calcul Scientifique en r\'egion Centre-Val de Loire (CCSC).

\bibliography{diracmilne_biblio}

\end{document}